\documentclass[traditabstract]{aa}  %longauth
\usepackage{graphicx}
\usepackage{textcomp}
\usepackage{lscape}
\usepackage{multicol}
\usepackage[varg]{txfonts}
\usepackage{mathrsfs}
\usepackage{amssymb}
\usepackage{natbib}
\usepackage[colorlinks, citecolor={blue}]{hyperref}
\usepackage[utf8]{inputenc}
\usepackage{color}
\usepackage{float}

\authorrunning{S.~Duarte~Puertas et al.,}
\titlerunning{MZ -- SFR in star-forming galaxies}

\begin{document} 

\title{Mass-Metallicity and Star Formation Rate in Galaxies: \\
a complex relation tuned to stellar age\thanks{Table~\ref{table:data} with the values of oxygen and nitrogen-to-oxygen abundance ratios for $\sim$195000 SDSS star-forming galaxies and related relevant data is only available in electronic form at the CDS via anonymous ftp to cdsarc.u-strasbg.fr (130.79.128.5) or via \texttt{http://cdsweb.u-strasbg.fr/cgi-bin/qcat?J/A+A/}.}}

\author{
S.~Duarte~Puertas\inst{1,2}
\and 
J.~M.~Vilchez\inst{2}
\and 
J.~Iglesias-P\'{a}ramo\inst{2}
\and
M.~Moll{\'a}\inst{3}
\and
E. P\'erez-Montero\inst{2}
\and
C.~Kehrig\inst{2}
\and
L.~S.~Pilyugin\inst{4,5}
\and
I.~A.~Zinchenko\inst{6,4}
}
\institute{
D\'epartement de Physique, de G\'enie Physique et d’Optique, Universit\'e Laval, and Centre de Recherche en Astrophysique du Qu\'ebec (CRAQ), Qu\'ebec, QC, G1V 0A6, Canada\label{inst1}
 \\ \email{salvador-manuel.duarte-puertas.1@ulaval.ca}
\and Instituto de Astrof\'{\i}sica de Andaluc\'{\i}a - CSIC, Glorieta de la Astronom\'{\i}a s.n., 18008 Granada, Spain\label{inst2}
\and CIEMAT, Avda. Complutense 40, 28040 Madrid, Spain\label{inst3}
\and Main Astronomical Observatory, National Academy of Sciences of Ukraine, 27 Akademika Zabolotnoho St., 03680, Kyiv, Ukraine\label{inst4}%03143
\and Institute of Theoretical Physics and Astronomy, Vilnius University, Sauletekio av. 3, 10257, Vilnius, Lithuania\label{inst5}
\and Faculty of Physics, Ludwig-Maximilians-Universit\"{a}t, Scheinerstr. 1, 81679 Munich, Germany\label{inst6}
}

\authorrunning{S. Duarte Puertas et at.}
\date{Received \today; accepted \today}

 \abstract 
% 5 {} token are mandatory
{In this work we study the stellar mass -- metallicity relation (MZR) of an extended sample of star-forming galaxies in the local Universe and its possible dependence with the star formation rate (SFR).}
{A sample of $\sim$195000 Sloan Digital Sky Survey (SDSS) star-forming galaxies has been selected up to z=0.22 with the aim of analysing the behaviour of the relation of MZR with respect to SFR and taking into account the age of their stellar populations.}
{For this sample we have obtained, for the first time, aperture corrected oxygen and nitrogen-to-oxygen abundances (O/H and N/O, respectively) and SFR using the empirical prescriptions from the Calar Alto Legacy Integral Field Area (CALIFA) survey. To perform this study we make use also of the stellar mass of the galaxies and the parameter $\rm D_n(4000)$ as a proxy of the age of the stellar population.}
{We derive a robust MZR locus, which is found to be fully consistent with the ``anchoring" points of a selected set of well studied nearby galaxies with a direct derivation of the chemical abundance. A complex relation between MZR and SFR across the whole range of galaxy mass and metallicity has been observed, where the slope changes seen in the O/H -- SFR plane present a pattern which seems to be tuned to the galaxies' stellar age, and therefore, stellar age has to be taken into account in the stellar mass -- metallicity -- SFR relation.}
{In order to provide an answer to the question of whether or not the MZR depends on the SFR it is essential to take into account the age of the stellar populations of galaxies. A strong dependence between the MZR and SFR is observed mainly for star-forming galaxies with strong SFR values and low $\rm D_n(4000)$. The youngest galaxies of our SDSS sample show the highest SFR measured for their stellar mass.}
\keywords{galaxies: general --
                galaxies: star forming --
                galaxies: formation --
                galaxies: evolution --
                galaxies: star formation rate --
                galaxies: aperture corrections --
                galaxies: metallicity
               }

\maketitle

%________________________________________________________________
%% 1. Introduction %%%%%%%%%%%%%%%%%%%%%%
\section{Introduction}
Since the pioneering work of \cite{1979A&A....80..155L}, a plethora of papers have studied the relation between metallicity and galaxy stellar mass, as well as with other fundamental parameters of galaxies \citep[e.g.][]{1992MNRAS.259..121V,2002ApJ...581.1019G,2004A&A...425..849P,2004ApJ...613..898T,2006ApJ...647..970L,2014ApJ...791..130Z,2017MNRAS.469.2121S,2019A&ARv..27....3M,2020ARA&A..5812120S}. In all of these works a clear correlation between total stellar mass $\rm (M_\star)$ and metallicity of galaxies was highlighted. The mass-metallicity relation (MZR) thus remains a key ingredient for our understanding of the formation and evolution of galaxies. Theoretical models predict a tight relation \citep[e.g.][]{2008A&A...486..711M,2011MNRAS.410.2203S,2011MNRAS.416.1354D,2013ApJ...770..155R,2017MNRAS.464.4866O,2017MNRAS.472.3354D,2018MNRAS.477L..16T} illustrating how galaxy evolution is driven by galaxy mass, modulated by gas accretion and outflows. 

The MZR reaches a maximum near saturation at the high mass end ($\rm \log(M_\star/M_\odot) \ge 10.5$), maintaining in a metallicity asymptotically close to the oxygen yield \citep[e.g.][]{2007MNRAS.376..353P}. The study of this relation for galaxies of different masses can be used to better ascertain if the MZR could be emerging from a local relationship within galaxies or rather represents a truly global scale relation, e.g. linked to galaxy mass. The overall shape of the MZR is believed to respond mainly to the action of galactic winds and enriched outflows in addition to metal astration driven by star formation and galaxy stellar mass, but other processes related to e.g. galaxy downsizing or massive gas accretion and star formation can be also involved \citep[e.g.][]{2008MNRAS.385.2181F}.

Star formation rate (SFR) was originally introduced as a secondary parameter of the MZR aiming at reducing the observed scatter, and in this way the so-called fundamental mass -- metallicity -- SFR relation was defined \citep[MZSFR; e.g.][]{2008ApJ...672L.107E,2010MNRAS.408.2115M,2010A&A...521L..53L}. In this fundamental relation, at a fixed galaxy stellar mass, SFR anti-correlates with metallicity, similarly to the predictions of some models about this possible secondary dependence of MZR with SFR \citep[e.g.][]{2013MNRAS.430.2891D}. An intense debate \citep[e.g.][]{2013A&A...554A..58S,2017MNRAS.469.2121S,2019MNRAS.484.3042S,2014ApJ...797..126S,2017ApJ...844...80B,2019A&A...627A..42C} has developed since then about the existence or not of this second dependence on SFR. The observations of a large number of galaxies obtained with integral field spectroscopy (IFS) in CALIFA \citep{2012A&A...538A...8S,2015A&A...576A.135G}, and MANGA \citep{2015ApJ...798....7B} surveys, or with integrated spectra by \cite{2013A&A...550A.115H}, reach to different conclusions which prevent to understand the exact role played by the SFR, and the possible evolution of MZSFR with redshift. 
 
One of the key differences among the above cited studies, which we face in the present work, is that the integrated galaxies analysis have been usually performed using bright emission lines measured on single aperture spectroscopy of galaxies (e.g. spectra SDSS, VVDS, zCOSMOS, DEIMOS) to derive the oxygen abundance (O/H) of the ionised gas, as a proxy of galaxy metallicity; and the SFR estimates for these galaxies were also derived from the same aperture spectral information, with present-day SFRs often computed from the luminosity of H$\rm \alpha$ emission. The aperture corrections are produced by incomplete or partial coverage of the observed galaxies (e.g. SDSS 3 arcsec fibre spectroscopy), and may produce critical effects, which must be corrected. IFS of a large sample of galaxies of CALIFA has provided the method to correct the emission lines by aperture and spatial sampling effects \citep{2014A&A...561A.129M,2016A&A...586A..22G,2016ApJ...826...71I}. It has been shown that aperture effects translate in clear flux deficits which, in the case of e.g. SFR, imply corrections of up to $\sim$0.6 dex \citep{2013A&A...553L...7I,2017A&A...599A..71D}; whereas model-based aperture corrections do not solve this problem. In the case of the emission-line fluxes the situation is not trivial, as they, and their corresponding extinction must be aperture corrected before deriving SFR or the different abundance ratios. In particular, differential extinction correction across the disks of galaxies can not be overlooked. In \cite{2016ApJ...826...71I} empirically CALIFA based aperture corrections are provided for the relevant lines.

The SFR of galaxies can be also derived using multiparametric fitting of their spectral energy distribution, using evolutionary population synthesis models and inverting population synthesis equations (with their own uncertainties). Recent spatially resolved studies using this last technique found a clear correlation between O/H and SFR using the IFS CALIFA and MANGA data \citep[e.g.][]{2019A&A...627A..42C}. However, according to \citet{2020ARA&A..5812120S}, whether or not a fundamental metallicity relation exists depends on the analysis of the data, i.e., with the same CALIFA data set, \cite{2014ApJ...797..126S} found a dependence between the sSFR and the metallicity not found by \cite{2013A&A...554A..58S}, who suggested that the MZSFR relation is an artifact of spectroscopy aperture bias. On the other hand, \citet{2019A&A...627A..42C} concluded that the method used to calculate the stellar mass of galaxies, the other key ingredient in the MZR, also affects the obtained results. To these considerations we need to add that other research using IFS studied the relationship between O/H and other parameters \citep[e.g.][]{2018ApJ...852...74B,2020MNRAS.492.2651B,2020arXiv200914211S} propose that there might be some dependence, even indirectly, between gas-phase metallicity and stellar age at local and global scales. Both questions are actually related, thus the study of a possible evolution of the zero point and slope of the MZR with cosmic time and its dependence on SFR, can shed light on how galaxies formed, were assembled and evolve, an also on the infall and galactic wind phases of their chemical evolution. The MZR has been extensively studied as a general scaling relation of star-forming galaxies in the local Universe \citep[e.g.][]{2010MNRAS.408.2115M,2010A&A...521L..53L,2016ApJ...823L..24K,2020MNRAS.491..944C}, as a function of galaxy environment \citep[e.g.][]{2009MNRAS.396.1257E,2011ApJ...734...32P,2012ApJ...749..133P,2014MNRAS.438..262P,2017MNRAS.465.1358P}, and in medium and high redshift surveys \citep[e.g.][]{2016MNRAS.458.1529B, 2016MNRAS.463.2002H,2017PASJ...69...44K}. Some evolution is expected for the shape or for the zero point of the MZR \citep[e.g.][]{2010A&A...519A..31L, 2013MNRAS.430.2680M, 2013MNRAS.432.1217P}; though, other works find no significant evolution for the MZSFR \citep[e.g.][]{2010MNRAS.408.2115M,2012MNRAS.421..262C}.

Besides the above uncertainties, we have the well known issue of metallicity derivation from spectroscopy in large samples, which has been discussed elsewhere \citep[e.g.][and references therein]{2017MNRAS.465.1384C}. Since temperature sensitive lines fluxes are not available for the huge majority of these objects (e.g. in SDSS), abundance calibrations are applied to derive their metallicity. Not all abundance calibrations appear equally reliable when they are compared with oxygen abundances derived from the direct method. In fact, recent abundance calibrations empirically calibrated provide an abundance which, statistically, is typically within 0.2 dex uncertainty of the direct value. In this respect, some support can be gained using complementary versions of the MZR relation, e.g. using stellar metallicity derived either from integrated young stellar populations \citep[e.g.][]{2005MNRAS.362...41G} or for individual massive stars \citep[e.g.][]{2016ApJ...830...64B}; or deriving O/H directly from stacked spectra of mass-grouped star-forming galaxies \citep[e.g.][]{2013ApJ...765..140A}; also, replacing O/H by nitrogen-to-oxygen abundance (N/O), an abundance ratio less dependent on electron temperature and a well known ``chemical clock" \citep[e.g.][]{1978MNRAS.185P..77E,2003A&A...397..487P,2006MNRAS.372.1069M} adding useful chemical evolution information. N/O also presents a well-known relation with stellar mass \citep{2009MNRAS.398..949P}, as for high metallicity nitrogen has mainly a secondary origin, and as its derivation does not depend on the excitation of the gas, the study of the mass vs N/O relation (MNOR) can thus be very useful to better understand the zero-point and the slope of the MZR. In addition the combined study of mass, SFR, and N/O can also be used to better discriminate the dependence of metallicity with SFR. In this work, the caveats mentioned above have been taken into account for the metallicity derivation and analysis, and the corresponding sanity checks have been performed when appropriate.
 
In summary, in order to carry out an in depth study of the relation between MZR and SFR, we must address the systematic effects involved, as discussed in \citet{2016ApJ...827...35T} and \cite{2019A&A...627A..42C}, as those associated to e.g. signal-to-noise ratio (S/N), aperture effects, or the metallicity indicators used, among others, which can affect the derivation of the MZR and any possible dependence of MZR with SFR. This is the main objective of this work. To do it, we benefit from the methodology and knowledge gained with the CALIFA survey and apply it here to minimise systematic effects. Hence we examine in detail the behaviour of the relation between MZR and SFR for a large and complete sample of SDSS star-forming galaxies \citep[see][for details]{2017A&A...599A..71D} using the total fluxes of their emission lines empirically corrected by extinction and aperture effects in a consistent manner. We will analyse the MZR and SFR relationships, studying in particular the role of the age, using the parameter $\rm D_n(4000)$ as a proxy, of the stellar populations of galaxies.

The structure of this paper is organised as follows: in Sect.~\ref{sec:2_data} we describe the data providing a description of the methodology used to select the sample(Sect.~\ref{sec:2_sample}). The methodology followed to derive all the parameters used in this work is presented in Sect.~\ref{sec:2_empirical_deriv}. Main results and discussion are given in Sect.~\ref{sec:3_discu}. Conclusions are given in Sect.~\ref{sec:4_conclusions}. Finally, supplementary material has been added in Appendix \ref{sec:sm}. Throughout the paper, we assume a Friedman-Robertson-Walker cosmology with $\Omega_{\Lambda 0}=0.7$, $\Omega_{\rm m 0}=0.3$, and $\rm H_0=70\,km\,s^{-1}\,Mpc^{-1}$. We use the \cite{2001MNRAS.322..231K} universal initial mass function (IMF).

%__________________________________________________________________
%% 2. Data and sample %%%%%%%%%%%%%%%%%%%%%%
\section{Data and sample}
\label{sec:2_data}

%__________________________________________________________________
%% 2.1 The sample of galaxies %%%%%%%%%%%%%%%%%%%%%%
\subsection{The sample of galaxies}
\label{sec:2_sample}
Our study is based on the catalogue of 209276 star-forming galaxies extracted from SDSS-DR12 \citep{2015ApJS..219...12A} and presented in \cite{2017A&A...599A..71D}. The galaxies span a redshift (z) and stellar mass ranges of $\rm 0.005 \leq z \leq 0.22$ and $\rm 8.5 \leq \log{(M_{\star}/M_{\odot})} \leq 11.5$, respectively. All the emission line fluxes (i.e. $\rm [OII] {\lambda\lambda\, 3727,\,3729}$, $\rm [OIII]{\lambda\, 5007}$, $\rm H\beta$, $\rm [NII] {\lambda\, 6584}$, and $\rm H\alpha$), 
$\rm \log{(M_{\star}/M_{\odot})}$, z, and $\rm D_n(4000)$\footnote{$\rm D_n(4000)$ corresponds to the narrow definition of the 4000 $\rm \AA$ break strength from \citealt{1999ApJ...527...54B} that can be considered a proxy of stellar population age.} used in this work have been taken from Max-Planck-Institut für Astrophysik and Johns Hopkins University (MPA-JHU) public catalogue\footnote{Available at \href{http://www.mpa-garching.mpg.de/SDSS/}{\texttt{http://www.mpa-garching.mpg.de/SDSS/}}.} \citep{2003MNRAS.341...33K,2004MNRAS.351.1151B,2004ApJ...613..898T,2007ApJS..173..267S}. From this catalogue we have selected a subset of 194353 SDSS star-forming galaxies according to the following criteria:
\begin{itemize}
\item{i)} $\rm z \geq 0.02$. Due to the SDSS spectral range (3800-9200 $\rm\AA$), this is the minimum redshift required to include and measure $\rm [OII] {\lambda\lambda\, 3727,\, 3729}$; this emission lines are necessary to derive the metallicity (see Sect.~\ref{sec:2_empirical_deriv}) from our spectra \citep{2012MNRAS.421.1624P,2013A&A...549A..25P}.
\item{ii)} A S/N $\geq$ 3 is imposed for all the line fluxes used to derive O/H and N/O.
\end{itemize}
For these objects, we will determine the oxygen and nitrogen-to-oxygen abundances, following the prescriptions explained in Sect. \ref{sec:2_empirical_deriv}.
For the SFR, we use aperture corrected measurements from the database of \citet{2017A&A...599A..71D}. 

%______________________________________________________________
%% 2.2 Empirical aperture correction and chemical abundances %%%%%%%%%%%%%%%%%%%%%%
\subsection{Empirical aperture correction and chemical abundances}
\label{sec:2_empirical_deriv}

It is well known that the 3 arcsec diameter fibres in SDSS only cover a limited region of galaxies in the low-z Universe (z $<$ 0.22). In order to obtain the total SFR for all the galaxies in our sample we have used the aperture corrected SFR values from \cite{2017A&A...599A..71D}, where a detailed description of the aperture correction for the $\rm H\alpha$ flux measurements in SDSS is presented. 

The aperture corrections for the $\rm H\alpha$ flux and $\rm \log([OIII] {\lambda\, 5007}/H\beta)$, $\rm \log([NII] {\lambda\, 6584}/H\alpha)$, and $\rm H\alpha/H_\beta$ flux ratios in our sample were performed using the prescriptions by \citet{2016ApJ...826...71I}. These corrections were derived for a sample of disk galaxies in the CALIFA survey, representative of different types and masses \citep{2016ApJ...826...71I}. The aperture correction recipe for $\rm [OII] {\lambda\lambda\, 3727,\, 3729}$ has been also provided to us \citetext{by \citeauthor[][private communication]{2016ApJ...826...71I}}. Conversely, as shown in \citet{2016MNRAS.461.3111B}, the $\rm D_n(4000)$ break strength index remains substantially constant across star-forming galaxies, given that their young stellar component appears well distributed through out all the galactic disk; hence no aperture correction was applied for it. In summary, for each galaxy in our sample, the line fluxes measured in the SDSS fibres (i.e. $\rm [OII]{\lambda\lambda\, 3727,\, 3729}$, $\rm [OIII]{\lambda\, 5007}$, $\rm H\beta$, $\rm [NII]{\lambda\, 6584}$, and $\rm H\alpha$) were corrected for extinction following \citet{2017A&A...599A..71D}, and then they were aperture-corrected applying the equations in Table 20 of \citet{2016ApJ...826...71I}. These emission line fluxes are necessary for the derivation of oxygen and nitrogen abundances, as explained.

In order to derive the metallicity of our sample galaxies, given that the faint temperature sensitive lines are not available, we must rely on bright lines abundance calibrations (including from [OII], [OIII] to [NII]). It is well known that some calibrations produce different absolute values of O/H; especial care has been exercised in this work in the selection of the abundance calibrations applied. We have used three different methods:
\begin{enumerate}
\item We first follow the study updated by \citet{2017MNRAS.465.1384C} for O/H derivation and have selected their calibration of O3N2 which presents the wider O/H prediction range with the smallest dispersion.
\item In addition, we have also adopted the empirical calibrations by \cite{2016MNRAS.457.3678P} for nitrogen (their eq. 13) and oxygen (their eqs. 4, 5) abundances. 
\end{enumerate}

% Table 1__________________________________________________________________
\begin{table*}
\begin{center}
\caption{Compiled data for our sample of star-forming galaxies. The whole table is available in electronic format.}
\label{table:data}
\begin{tabular}{cccccc}
\hline
\hline \\[-2ex]
 (1) & (2) & (3) & (4) & (5) & (6) \\
specObjID & 12+log(O/H) & log(N/O) & $\rm \log({M_{\star}})$ & log(SFR) & $\rm D_{n}(4000)$ \\
                 &                               &                          &         [$\rm M_{\odot}$] & [$\rm M_{\odot}\,yr^{-1}$] &   \\
                 &          & & & & \\
\hline
1399656635961468928 & 8.44$\pm$0.07 & -1.23$\pm$0.11 & 9.98$^{+0.15}_{-0.12}$ & 0.86$\pm$0.02 & 1.45$\pm$0.17 \\[0.5ex]
1302792405518411776 & 8.16$\pm$0.07 & -1.38$\pm$0.06 & 9.95$^{+0.32}_{-0.47}$ & -0.28$\pm$0.01 & 1.21$\pm$0.03 \\[0.5ex]
1302813296239339520 & 8.38$\pm$0.02 & -1.17$\pm$0.02 & 10.23$^{+0.33}_{-0.48}$ & 0.32$\pm$0.01 & 1.19$\pm$0.03 \\[0.5ex]
1398489229004138496 & 8.53$\pm$0.02 & -0.79$\pm$0.05 & 10.64$^{+0.11}_{-0.08}$ & 1.45$\pm$0.01 & 1.20$\pm$0.03 \\[0.5ex]
1399648389624260608 & 8.40$\pm$0.01 & -1.10$\pm$0.02 & 9.51$^{+0.13}_{-0.08}$ & 0.34$\pm$0.01 & 1.12$\pm$0.05 \\[0.5ex]
... & ... & ... & ... & ... & ... \\
\end{tabular}
\end{center}
\end{table*}
%__________________________________________________________________

% Figure 1__________________________________________________________________
\begin{figure*}
    \centering
    \includegraphics[width=\columnwidth]{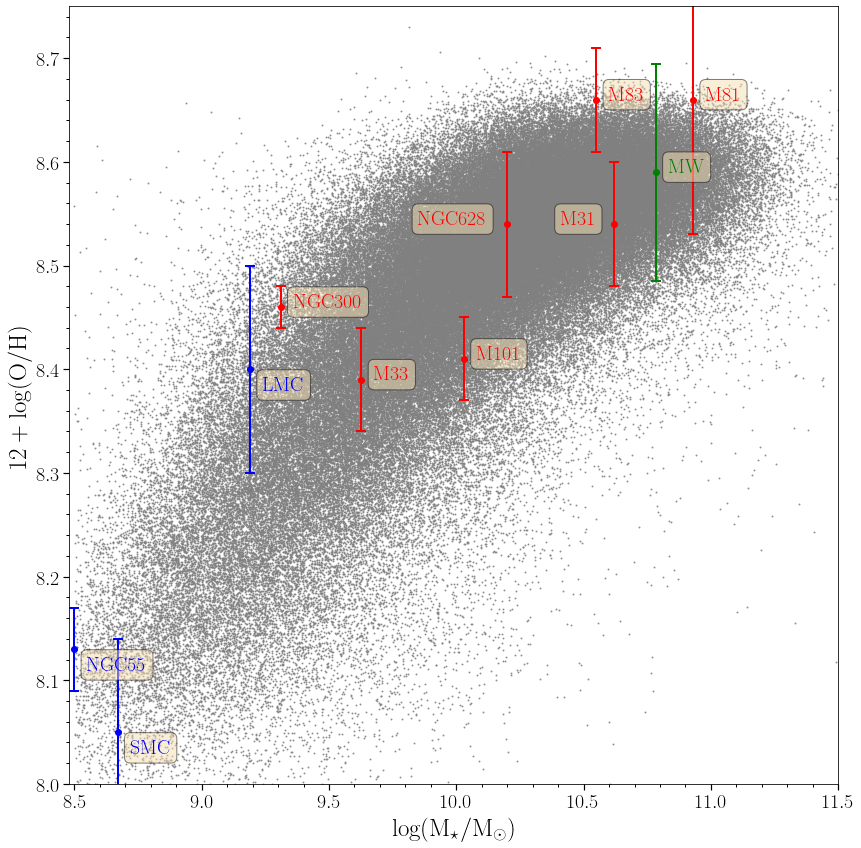}
    \includegraphics[width=\columnwidth]{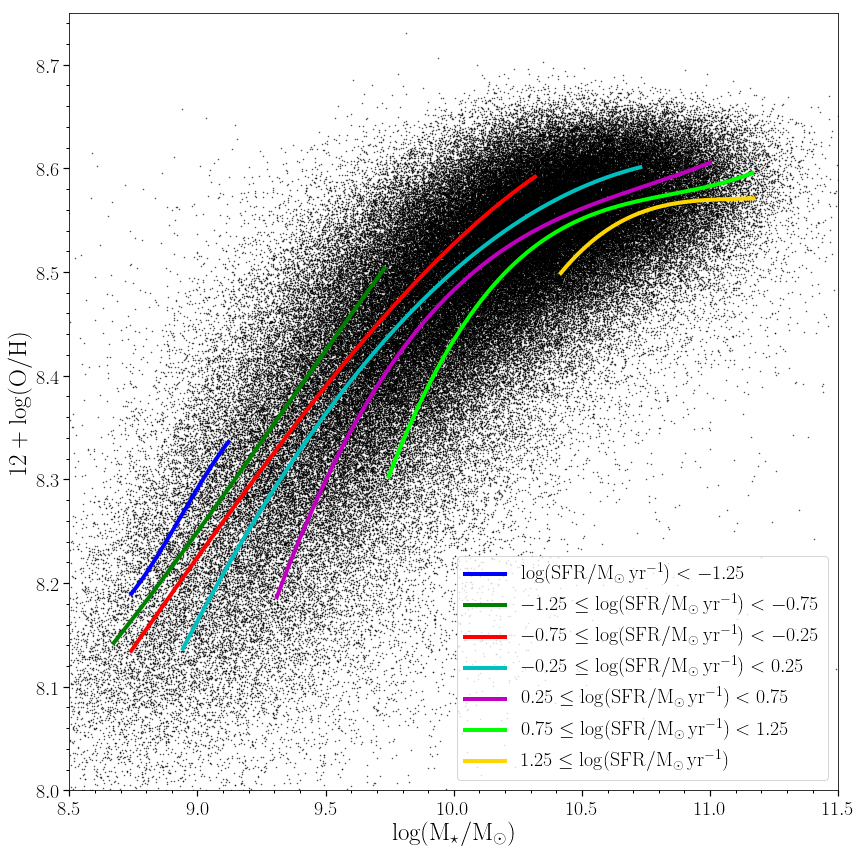}
    \caption{a) Aperture corrected MZR. Star-forming galaxies of our sample (grey dots), as well as the Milky Way (green mark) and other nearby galaxies (labelled; coloured points) with direct (from electron temperature) oxygen abundances are shown. Blue or red colour indicate galaxies with a flat or non-zero radial oxygen gradient, respectively. \protect\footnotesize \protect\emph{References as follows: galaxy name, O/H reference, stellar mass reference; SMC, LMC, and M81, \protect\cite{2016ApJ...830...64B}, \protect\cite{2012ApJ...747...15K}; NGC 55, \protect\cite{2017MNRAS.464..739M}, \protect\cite{2006ApJ...647..970L}; MW, \protect\cite{2017A&A...597A..84F,2021MNRAS.502..225A}, \protect\cite{2015ApJ...806...96L}; M83, \protect\cite{2016ApJ...830...64B,2019ApJ...872..116H}, \protect\cite{2016A&A...585A..20K}; M33 and NGC 300, \protect\cite{2016ApJ...830...64B}; \protect\cite{2016A&A...585A..20K}; M 31, \protect\cite{2012MNRAS.427.1463Z}, \protect\cite{2011ApJ...733L..47F}; M 101, \protect\cite{2016ApJ...830....4C}, \protect\cite{2011ApJ...738...89S}; NGC 628, \protect\cite{2015ApJ...806...16B}, \protect\cite{2014MNRAS.445..899C} (see also \citealt{2019MNRAS.483.4968V})}. b) Relation between 12+log(O/H) and $\rm M_\star$ for SDSS star-forming galaxies, grey points. Overplotted blue, green, red, cyan, magenta, lime, and yellow solid lines represent the fits to the running median of the SFR of the galaxies, calculated in bins of 1000 galaxy points, corresponding to seven SFR intervals as indicated.}
    \label{fig:1_MZRSFR} 
\end{figure*}
%__________________________________________________________________
All these calibrations are empirical, i.e. they are calibrated against direct abundances calculated with electron temperature measurements. Finally, we have used the ones given by the code \textsc{HII-CHI-mistry} \citep{2014MNRAS.441.2663P}, based on photoionization models, and which reproduce chemical abundances consistent with their corresponding direct derivations\footnote{A known feature is that most photoionisation model abundances are typically overestimated; while \textsc{HII-CHI-mistry} abundances appear consistent with direct derivations}. For the sake of consistency to compare with previous and other works performed including spectroscopy of distant galaxies, oxygen and nitrogen-to-oxygen abundances for our sample galaxies have been derived with the three methods mentioned above once they are free from aperture effects (given that the emission line fluxes used were all aperture corrected), corresponding to the entire galaxy. We have also derived the value of oxygen and nitrogen-to-oxygen abundances in fibre and we have obtained that the values derived in the fibre, for O/H and N/O, are systematically higher than those aperture corrected. The mean oxygen abundance difference between the fibre and the value corrected for aperture, $\Delta$(12+log(O/H)), is 0.04 dex (the standard deviation is 0.01 dex) for our sample of star-forming galaxies. In the case of the mean nitrogen-to-oxygen abundance difference, $\rm \Delta$(log(N/O)), we have found a value of 0.06 dex (the standard deviation is 0.01 dex). All the abundances obtained for each object resulted consistent within the errors, and overall, the three abundance outputs from the calibrations selected (and values in the fibre) are statistically consistent to within $\sim$0.15 dex; with the higher consistency and lowest errors being achieved especially for the nitrogen to oxygen abundance ratio. Taking this fact into account, in this work we adopt for each galaxy as representative the O/H and N/O abundances obtained applying \citet{2016MNRAS.457.3678P}. As a result, we have compiled Table~\ref{table:data} where we show a sample of the online table of 194353 star-forming galaxies, where their fluxes have been aperture-corrected, for which we have the spectroscopic identifier from SDSS (specObjID), oxygen, as 12+log(O/H), and nitrogen-to-oxygen abundance ratio, as log(N/O), abundances, stellar mass, as $\rm \log{(M_{\star}/M_{\odot})}$, the SFR (in $\rm M_{\odot}\,yr^{-1}$ units), and the parameter $\rm D_n(4000)$ that we use as the stellar population age indicator. Table~\ref{table:data} also shows the uncertainties for each property considered.\footnote{The uncertainties of $\rm \log{(M_{\star}/M_{\odot})}$ and $\rm D_n(4000)$ have been taken from MPA-JHU.}

%__________________________________________________________________
%% 3. Discussion %%%%%%%%%%%%%%%%%%%%%%
\section{Discussion}
\label{sec:3_discu}

In Fig.~\ref{fig:1_MZRSFR}, panel a, we show the oxygen abundance, $\rm 12+log(O/H)$, versus total stellar mass for the 194353 galaxies in our sample, which clearly reproduce the MZR relation. A further quality check of our MZR derivation can be done comparing it with the oxygen abundance and total stellar mass\footnote{NGC 55, NGC 300, and M 33 stellar masses transformed from \cite{2003ApJ...586L.133C} to \cite{2001MNRAS.322..231K} IMF dividing by 0.943 \citep{2010MNRAS.408.2115M}.} for the Milky Way\footnote{An effective radius face value of 5.9 kpc was assumed for the Milky Way \citep{1997ApJ...483..103S,2009A&A...505..497Y}.} and ten well known nearby galaxies for which precise abundance values of their H{\sc ii} regions have been derived using the direct method from available electron temperature measurements. These nearby galaxies show well defined spatially resolved radial abundance gradients, thus their representative abundance, corresponding to the integrated galaxy flux can be easily derived using the IFS measurements in the CALIFA survey. To do so we use the O/H measured in the characteristic radius (0.4$\times$ R$_{25}$, basically similar to the effective radius of the galaxy, e.g. \citealt{2013A&A...554A..58S}), the most suitable representative value of the whole galaxy abundance. According to \cite{2013A&A...554A..58S}, for galaxies presenting $\rm 12+log(O/H)$ abundance at the effective radius, $\rm 12+log(O/H)_{Reff}$ $\le$ 8.6 dex, the mean difference between $\rm 12+log(O/H)$ of the integrated galaxy flux and $\rm 12+log(O/H)_{Reff}$ is $\rm \sim$-0.03 dex; whereas for galaxies with $\rm 12+log(O/H)_{Reff}$ above 8.6 dex, this difference amounts to $\rm \sim$0.06 dex. We have applied this conversion to calculate the representative $\rm 12+log(O/H)$ abundances of our selected nearby galaxies; these objects are also shown in Fig.~\ref{fig:1_MZRSFR} panel a. It can be seen that the selected sample of nearby galaxies and our sample of SDSS star-forming galaxies show an excellent agreement delineating the local universe MZR, free from aperture effects, derived in this work. As mentioned in Section~\ref{sec:2_empirical_deriv}, we have derived (and compared) the O/H values using the empirical calibrations from \citet{2016MNRAS.457.3678P, 2009MNRAS.398..949P}, the direct method by using the electron temperature for objects with auroral line measurements, and the theoretical photoionisation models (HII-CHI-mistry) from \citet{2014MNRAS.441.2663P}. All these methods lead to overall consistent abundances -within the errors- and the differences between them have already been discussed. In Fig.~\ref{fig:A5_OH_te} it can be seen that all galaxies where the electron temperature has been calculated show a high consistency between the O/H from the direct method and the corresponding values obtained in this work using empirical calibrators. Therefore we consider that the oxygen abundances used are precisely estimated and better than others in the literature.

As we also want to see the role of the SFR in the MZR relation, we show in panel b) of Fig.~\ref{fig:1_MZRSFR} the diagram for the same MZR as in panel a), but overplotting the lines representing the fits to the running medians of seven different SFR intervals. The strong correlation between oxygen abundance and galaxy stellar mass, shown in the MZR plot, is also evident in all the shown SFR lines separately. We can see also how these iso-SFR loci tend to saturate at higher abundances and level off near solar metallicity\footnote{Solar metallicity: $\rm 12+\log(O/H)_{\odot}=8.69$ \citep{2009ARA&A..47..481A}.}. We may also see that for a fixed $\rm \log(M_\star/M_\odot)$, galaxies with a higher metallicity present lower SFR than lower metallicity galaxies, in line with previous findings \citep[e.g.][]{2010MNRAS.408.2115M}.

\begin{figure*}
    \centering
        \includegraphics[width=.48\textwidth]{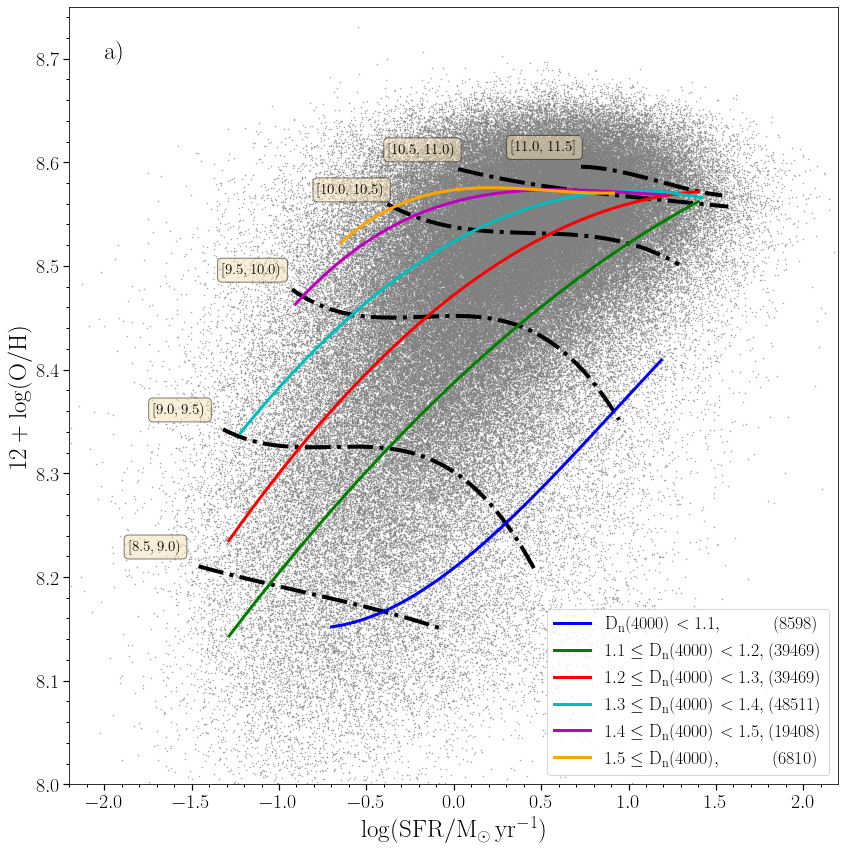}
        \includegraphics[width=.49\textwidth]{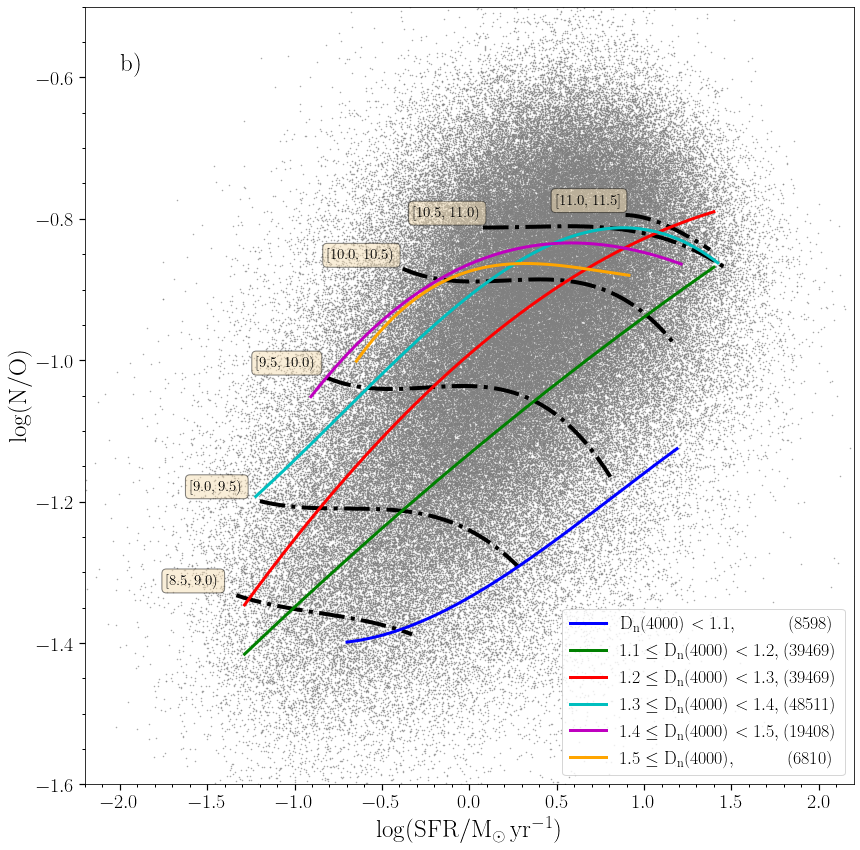}
\caption{a) Relation between the 12+log(O/H) and SFR corrected for aperture for star-forming galaxies. Blue, green, red, cyan, magenta, and yellow solid lines represent the fits to the running median for bins of 1000 objects in six $\rm D_{n}(4000)$ bins. The number of star-forming galaxies in each $\rm D_{n}(4000)$ bin appear on the legend. Black dash doted lines show the 12+log(O/H) {\sl vs.} SFR values of the running median fits for bins of 1000 objects in six $\rm \log{(M_{\star}/M_{\odot})}$ bins (each box shows the considered $\rm \log{(M_{\star}/M_{\odot})}$ range). b) Relation between the log(N/O) and SFR for star-forming galaxies (colours and numbers of star-forming galaxies in each $\rm D_{n}(4000)$ bins as in a).}
\label{fig:3_OHNOSFR_d4000}
\end{figure*}

We now analyse the relationship between the abundances and the SFR. In panel a) of Fig.~\ref{fig:3_OHNOSFR_d4000}, we show our aperture corrected oxygen abundances, as $\rm 12+log(O/H)$, versus the SFR for the galaxy sample (grey points) following the same statistical procedure as in the previous figure. We computed the running medians for six intervals of $\rm \log{(M_{\star}/M_{\odot})}$ with $\rm \Delta \log{(M_{\star}/M_{\odot})}=0.5$\,dex each (see also Figure~\ref{fig:A1_MZRSFR_binn_fit}). Their corresponding fits are plotted by black-dotted lines. The total covered range in galaxy stellar mass is $\rm 8.5 \leq \log{(M_{\star}/M_{\odot})}\leq 11.5$ in both panels. The overlaid colour dashed lines show the fits to the running median of the $\rm D_n(4000)$ index of the sample galaxies, calculated for six intervals from $\rm D_n(4000)<1.1$ to $\rm D_n(4000) \sim 1.5$, with $\rm \Delta D_n(4000)=0.1$ dex each, as indicated in the plot; in parenthesis the total number of points per interval is quoted. We can see a clear overall trend between 12 + log(O/H) against SFR in our sample (grey points). In panel b) of the same Fig.~\ref{fig:3_OHNOSFR_d4000} we similarly show the aperture corrected $\rm log(N/O)$ versus SFR. There, we have again the $\rm log(N/O)$ versus SFR considering the same galaxy mass and $\rm D_n(4000)$ bins, and fits to the running median. A similar strong correlation is also present between these two properties. A wide range in $\rm log(N/O)$ is covered by our SDSS galaxy sample, going from over-solar\footnote{Solar log(N/O)$_\odot=-0.86$ \citep{2009ARA&A..47..481A,2011MNRAS.412.1367T}.} values to the level typical of low metallicity dwarf galaxies ($\rm log(N/O) \sim$ -1.5). All the lines shown in Fig.~\ref{fig:3_OHNOSFR_d4000} (i.e. black for stellar mass and coloured for $\rm D_n(4000)$) are representing the real statistical (median) loci of the galaxy points and they are the result of the fit to the statistical distribution of the properties considered (i.e. $\rm D_n(4000)$, stellar mass, and O/H) for the SDSS star-forming galaxies sample used in this work.

% Figure 2__________________________________________________________________
%__________________________________________________________________

\begin{figure}
\includegraphics[width=\columnwidth]{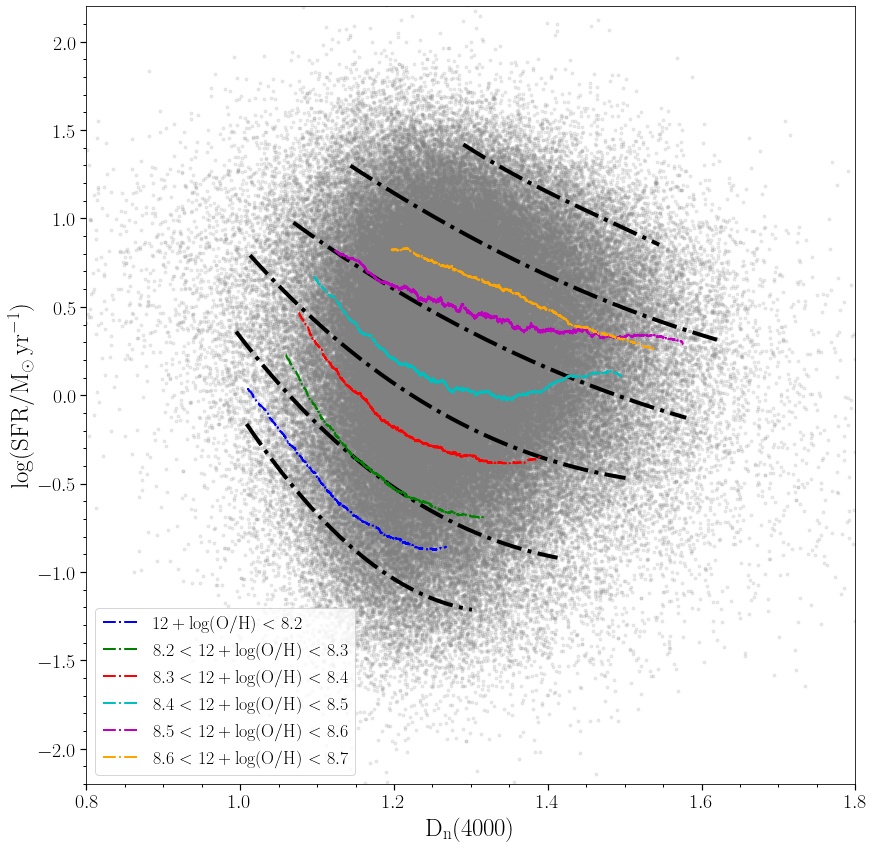}
\caption{Relation between the SFR and $\rm D_{n}(4000)$ for our sample of star-forming galaxies. Blue, green, red, cyan, magenta, and yellow solid lines represent the running median for bins of 2000 objects in six O/H bins. Black dash doted lines show the fit to the running median for bins of 1000 objects in six $\rm \log{(M_{\star}/M_{\odot})}$, bins, from low stellar mass (8.5, lower part of the figure) to high stellar mass (11.5, higher part of the figure).}
\label{fig: SFR_age}
\end{figure}

It is important to bear in mind that $\rm D_{n}(4000)$ depends on stellar population age, presenting lowest values for the youngest stellar populations, while the older ones show higher values (\citealt{1999ApJ...527...54B}). In panels a) and b) of Fig.~\ref{fig:3_OHNOSFR_d4000} we can see how for a fixed $\rm \log{(M_{\star}/M_{\odot})}$, galaxies with lower SFR present systematically higher values of $\rm D_{n}(4000)$, suggesting that they host older stellar populations. For $\rm D_{n}(4000) > 1.2$, we can see that the lines corresponding to the most massive galaxies ($\rm \log{(M_{\star}/M_{\odot})} > 10.5$) tend to saturate, and seem to flatten above $\rm 12+\log{(O/H)} \ge $ 8.55 and $\rm \log{(N/O)} \ge $ -0.9, irrespective of their SFR. For nearly the entire mass range of the sample ($\rm \log{(M_{\star}/M_{\odot})} \le 11$), oxygen abundance, $\rm 12+\log{(O/H)}$, shows a nearly flat or very mild decrease with respect to SFR along the black-dotted lines. Although the usual assumption is that SFR depends on age, showing an anticorrelation between stellar age and SFR when the stellar mass is taken into account, actually we show in this plot that both effects are disentangled. In Fig.~\ref{fig: SFR_age} we have plotted the SFR as a function of $\rm D_{n}(4000)$ to see more clearly this SFR-age effect. In this figure, we see that at face value there is a correlation between high SFR and high $\rm D_{n}(4000)$ values. Then, we overplotted the mass isolines and O/H isolines for six stellar mass bins and six O/H ranges, respectively, for our sample of SDSS star-forming galaxies. We can see how: i) at a fixed SFR the corresponding entire range of $\rm D_{n}(4000)$ is covered; ii) at fixed stellar mass, the lower the SFR, the higher the $\rm D_{n}(4000)$; and iii) at fixed O/H we see that the relationship is more complex: from $\rm D_{n}(4000) \sim 1.2$ towards the right, for O/H $<$ 8.5 the SFR stops decreasing with $\rm D_{n}(4000)$. For the left region, however, the anticorrelation between SFR and $\rm D_{n}(4000)$ is very strong. We consider this relationship to be relevant and deserving of further exploration with our sample of star-forming galaxies.

% Figure 3__________________________________________________________________

Depending on the stellar mass range, there is a value of SFR for which a change in O/H is observed. This SFR value varies depending on the stellar mass range considered. The iso-$\rm D_{n}(4000)$ line between [1.1-1.2) accompanies the change in the O/H -- SFR relation depending on the stellar mass, which suggests that this line determines a statistical value for stellar mass, SFR, and age.\footnote{This iso-$\rm D_{n}(4000)$ line around 1.1-1.2, as the others, is derived as the statistical median of galaxies with $\rm D_{n}(4000)$ values between 1.1 and 1.2. It is therefore the geometric locus of SDSS star-forming galaxies within these $\rm D_{n}(4000)$ values.} The data do not allow us to say the exactly age where the slope of these iso-mass lines change in the O/H-SFR diagram, but our results allow us to say that it must occur around that $\rm D_{n}(4000)$ value (i.e. in the green line). Therefore, in order to address this point, we have proceeded as follows:

 \begin{itemize}
 \item We derived $\rm \frac{dO/H}{dSFR} \Bigr\rvert_{M_\star}$, i.e. the derivative of O/H versus SFR at constant stellar mass, along all iso-mass lines below the range 10 $\leq$ log(M$_\star$/M$_\odot$) $<$ 10.5; (that means for four iso-mass lines).
 \item We obtain the value of log(SFR) for which the derivative presents the net measurable change, $\rm log(SFR_{deri})$, with a minimum value of $-0.002$, the half of the value obtained for the iso-mass line of the range 8.5 $\leq$ log(M$_\star$/M$_\odot$) $<$ 9.0; That is the point where the slope starts to be steeper (and negative) than this value.
 \item The iso-mass line corresponding to this lowest mass range plotted shows a nearly constant derivative, approximately a straight line in the plot without any apparent change in its slope; therefore, any change in the derivative of this iso-mass line should have occurred for $\rm \log{(SFR)} < -1.5$ (alternatively, there exists no such change in the derivative of the sub-sample of SDSS galaxies in these stellar mass and SFR range or it is small and we cannot statistically ``see'' it)
 \item We determine $\rm \log(SFR_{cut})$ as the SFR value at which each iso-$\rm D_{n}(4000)$ line cut each iso-mass line in the O/H vs. SFR diagram in Fig.~\ref{fig:3_OHNOSFR_d4000}.
\item Then we compute the difference between $\rm \log{(SFR_{cut})}$ and $\rm \log{(SFR_{deri})}$, as $\rm \Delta\log{(SFR)}=\log{(SFR_{cut})}-\log{(SFR_{deri})}$, for each iso-mass line. Since the iso-$\rm D_{n}(4000)$ line between 1.2 and 1.3 does not intersect the $\rm 8.5 \le log(M_\star/M_\odot) < 9.0$ iso-mass line in the O/H {\sl vs} SFR diagram, we assume the position in this diagram at which it would intersect by extrapolating slightly the iso-$\rm D_{n}(4000)$ line for the sake of this test.
\end{itemize}

% Figure 4__________________________________________________________________
\begin{figure}
\includegraphics[width=\columnwidth]{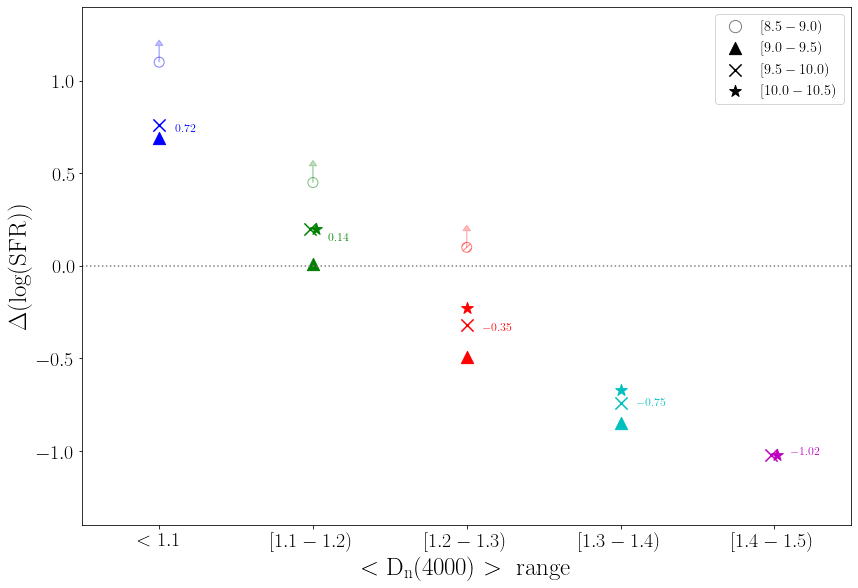}
\caption{Relation between $\rm \Delta\log{(SFR)}$ and the iso-$\rm D_{n}(4000)$ ranges. The symbols are colour-coded depending on the $\rm D_{n}(4000)$ range and according to Fig.~\ref{fig:3_OHNOSFR_d4000}, being $\rm D_{n}(4000) < 1.1$ (blue), $\rm 1.1 \le D_{n}(4000) < 1.2$ (green), $\rm 1.2 \le D_{n}(4000) < 1.3$ (red), $\rm 1.3 \le D_{n}(4000) < 1.4$ (cyan), and $\rm 1.4 \le D_{n}(4000) < 1.5$ (magenta). The symbol type is related to $\rm \rm \log{(M_{\star}/M_{\odot})}$ range considered, as labelled. For the stellar mass range 8.5-9.0 (circles) only a lower limit, if any, for $\rm \Delta\log{(SFR)}$ could be assumed from the present SDSS sample, leading to $\rm \log{(SFR)} < -1.5$. In the case of the iso-$\rm D_{n}(4000)$ line $\rm 1.2 \le D_{n}(4000) < 1.3 $, the value of the intersect position has been derived extrapolating slightly this iso-$\rm D_{n}(4000)$. The mean value of $\rm \Delta\log{(SFR)}$ for each iso-$\rm D_{n}(4000)$, considering all the iso-masses, is shown in the coloured numbers (only lower limits of $\rm \Delta\log{(SFR)}$ for the $\rm 8.5 \le log(M_\star/M_\odot) < 9.0$ mass range are not considered). The black dotted line indicates $\rm \Delta\log{(SFR)}=0$.}
\label{fig: 3_DeltaSFR}
\end{figure}

We show in Fig. \ref{fig: 3_DeltaSFR} the resulting $\rm \Delta\log{(SFR)}$ for the different stellar mass ranges we have considered, as a function of the iso-$\rm D_{n}(4000)$ ages. The symbols are coloured-code according to the colour of the iso-$\rm D_{n}(4000)$ lines of Fig.~\ref{fig:3_OHNOSFR_d4000}. We see as the green points show the closest values to $\rm \Delta\log{(SFR)}=0$. The mean value of $\rm \Delta\log{(SFR)}$= 0.72, 0.14, -0.35, -0.75, and -1.02 for galaxies in the $\rm D_{n}(4000)$ ranges $< 1.1$, [1.1-1.2), [1.2-1.3), [1.3-1.4) and [1.4-1.5), respectively. The difference is proportionally larger as we move away from the iso-$\rm D_{n}(4000)$ with values between 1.1 and 1.2 (green line), i.e., this iso-$\rm D_{n}(4000)$ line is the one with values of $\rm \Delta\log{(SFR)}$ closest to zero in all iso-masses considered in this figure. From this, we may consider that the iso-$\rm D_{n}(4000)$ line for $\rm 1.1 \le D_{n}(4000) < 1.2$ (green line in Fig.~\ref{fig:3_OHNOSFR_d4000}) represents the best approximation to the locus where the change of the relation between oxygen abundance and SFR is observed, for each stellar mass range and given SFR. Before this line, for $\rm D_{n}(4000)$ larger than 1.2, the slope of each O/H-SFR iso-mass line may be considered flat as its absolute value is smaller than 0.002.

This way, for all galaxies with stellar mass $\rm \log{(M_{\star}/M_{\odot})} \le 10.5$, the green iso-$\rm D_{n}(4000)$ line defines, {\it de facto}, an effective {\it isochrone}, which marks the locus of the sudden change in the derivative of oxygen abundance against SFR in the sample galaxies. This plot may illustrate why the case for a universal negative relation between oxygen abundance and SFR is still controversial for some samples of galaxies. The exact region in which the negative slope of the metallicity -- SFR relation may hold varies for each stellar mass range; a negative dependence of metallicity on SFR could be seen for a given galaxy mass range but only for those galaxies hosting the youngest stellar population (i.e. $\rm D_{n}(4000) \le 1.1$). A similar behaviour can be observed in panel b) for $\rm \log{(N/O)}$. Again for galaxies with $\rm \log{(M_{\star}/M_{\odot})} \lesssim 11$ there is a mild, if any, dependence of $\rm \log{(N/O)}$ on SFR, except when the green iso-$\rm D_{n}(4000)$ line is reached. Beyond this point, moving towards larger SFR for a fixed stellar mass (i.e. along dash-dotted lines), a strong negative dependence of $\rm \Delta \log{(N/O)}/\Delta \log{SFR}$ is evident. The value of $\rm \log{(N/O)}$ at which these changes in the slope can be seen from the typical values in dwarf galaxies up to the solar $\rm \log{(N/O)}_{\sun}$. The similar behaviour exhibited by the oxygen abundance and the nitrogen-to-oxygen abundance ratio against SFR, shown in both panels of Fig.~\ref{fig:3_OHNOSFR_d4000}, respectively, can give us some hints to understand the chemical evolution of our sample galaxies and the origin underlying of the MZRSFR relationship. 

We have found respective loci in the planes $\rm 12+\log{(O/H)}$ -- SFR and $\rm \log{(N/O)}$ -- SFR delimiting two broad regions domains in these diagrams for the sample galaxies, in one of them the sample galaxies show a clear dependence of metallicity on the SFR, whereas in the corresponding complementary region no dependence is seen, along each galaxy stellar mass sequence. These loci appear well represented on each plot by the line corresponding to galaxies with $\rm D_{n}(4000) < 1.2$, for which a mean stellar age younger than 150 Myr is expected \citep{2006MNRAS.370..721M}. Along these two loci lines, it seems that the precise oxygen abundance (N/O ratio) and SFR, for each galaxy stellar mass (i.e. specific SFR) appear to be tuned\footnote{By tuned, we refer that the stellar age accompanies and it has to be taken into account in the mass-metallicity-SFR relation.}, defining, {\it de facto}, the {\it effective isochrone} corresponding to the average stellar population for each galaxy. This applies only to galaxies with $\rm \log{(M_{\star}/M_{\odot})} \lesssim 10.5$; for more massive galaxies we do not observe this behaviour in our sample, since the lines of the $\rm D_{n}(4000)$ index converge at high metallicity (near solar to oversolar), and O/H and N/O do not show here observable decrease, suggesting an evolution without metallicity loss and mainly driven by galaxy mass.

\begin{figure*}
    \centering
    %\begin{subfigure}[b]{0.49\textwidth}
        \includegraphics[width=.46\textwidth]{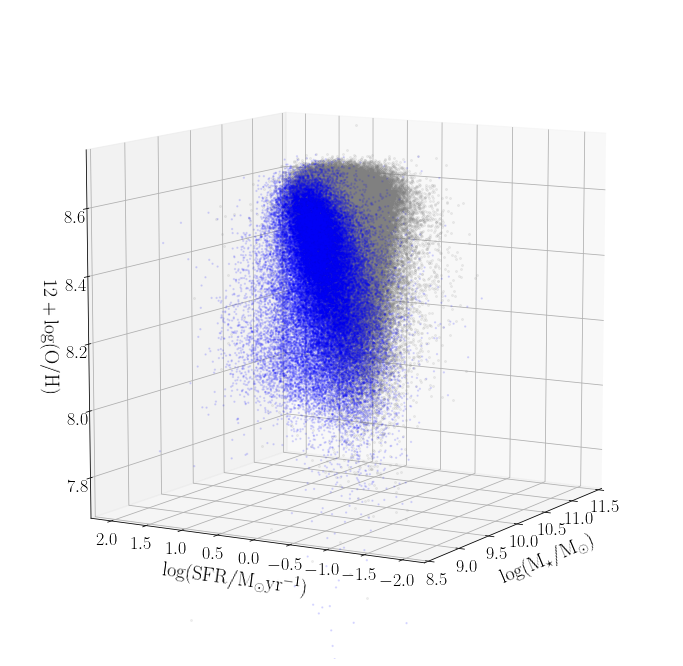}
    %\end{subfigure}%
    %\begin{subfigure}[b]{0.49\textwidth}
        \includegraphics[width=.50\textwidth]{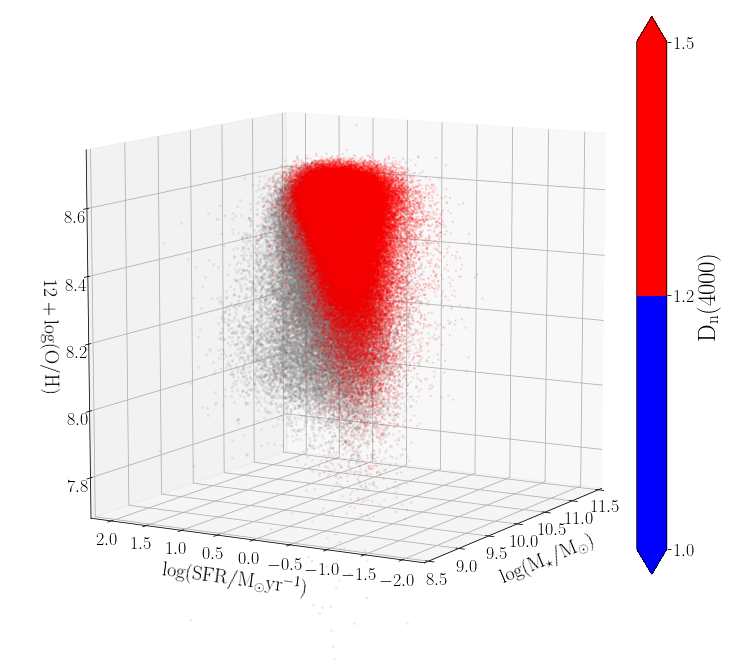}
    %\end{subfigure}
\caption{MZSFR for star-forming galaxies corrected for aperture. Colour-coded star-forming galaxies with $\rm D_{n}(4000) < 1.2$ (blue in panel a, represented as grey in panel b) and $\rm D_{n}(4000) \ge 1.2$ (red in panel b, represented as grey in panel a). Note the scale of the SFR axis decreases towards the right in the figures.}
\label{fig:4_FMZR_d4000}
\end{figure*}

\begin{figure*}
    \centering
    \includegraphics[width=\textwidth]{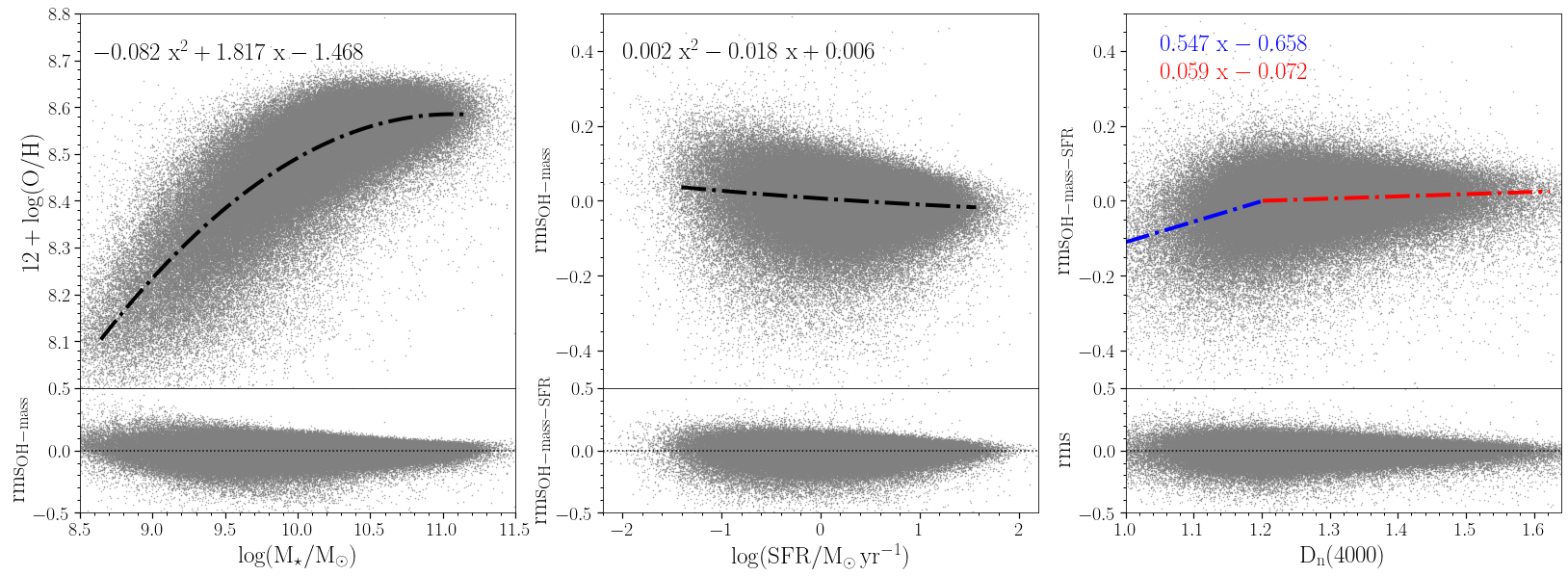}
    \caption{a) MZSFR for our sample of galaxies, overplotted a line with a polynomial fit to these results. In the bottom the residuals data-fit, $\rm OH-OH_{fit,M}$ are represented as a function of the stellar mass; b) These residuals of $\rm OH-OH_{fit,M}$ as a function of the SFR with a minimum straight line fitting the points. The residual of this fit are represented at the bottom as a function of SFR: c) The residual of the later fit, $\rm OH-OH_{fit,M}-Res_{fit,SFR}$ as a function of the parameter $\rm D_{n}(4000)$. In the upper panels, the correlation coefficients for each fit are shown in the top right.}
\label{fig: residuals}
\end{figure*}

In Fig.~\ref{fig:4_FMZR_d4000} we show the MZSFR relation for the star-forming galaxies with $\rm D_{n}(4000) < 1.2$ (panel a) and $\rm D_{n}(4000) \ge 1.2$ (panel b). From this figure, it can be seen that galaxies with young and old stellar populations are located in different domains in the MZSFR three dimensional diagram. As expected, galaxies with younger stellar populations ($\rm D_{n}(4000) < 1.2$, represented as blue dots in the figure) are located in the zone where SFR is higher, at a fixed 12+log(O/H). In contrast, galaxies with older stellar populations ($\rm D_{n}(4000) \ge 1.2$, represented as red dots in the figure) fall in the area where SFR is lower, at fixed 12+log(O/H). This plot has, however, a large dispersion, and therefore, we have analysed this possible dependence of the OH on SFR and on the stellar age from other point of view by using the classical method of representing residuals of a correlation as a function of other parameters. Thus, we plot three panels in Fig.~\ref{fig: residuals}. In the first one at the left, the MZR relation is drawn with a polynomial fit overplotted and, at the bottom, the residuals of this relation. In the middle panel, these same residuals are plotted as a function of the SFR, with a least-square straight line overplotted and the corresponding residuals at the bottom. These residuals, within $\pm 0.2$ dex, are represented as a function of the $\rm D_{n}(4000)$ parameter in the right panel, where again a correlation appears. This way, a third parameter, the age, besides the stellar mass and the SFR, arises clearly as something to take into account in the analysis of the MZR-SFR. In fact, in this last panel, a fit with two straight lines may be also acceptable since, as we claim before, for $\rm D_{n}(4000) > 1.2$ the correlation is almost flat, while a strong slope would appear in the younger objects with $\rm D_{n}(4000) <$ 1.2.

% Figure 5__________________________________________________________________
%__________________________________________________________________
% Figure 6__________________________________________________________________
%__________________________________________________________________

\begin{figure*}
\centering
    \includegraphics[width=.48\textwidth]{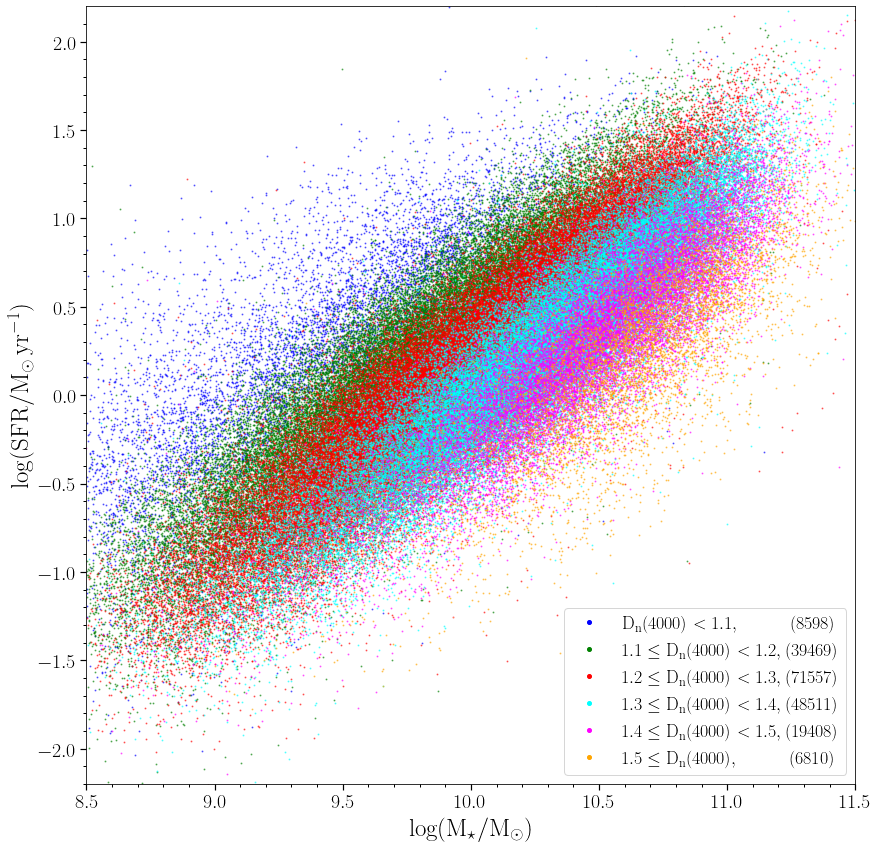}
\hfill
    \includegraphics[width=.48\textwidth]{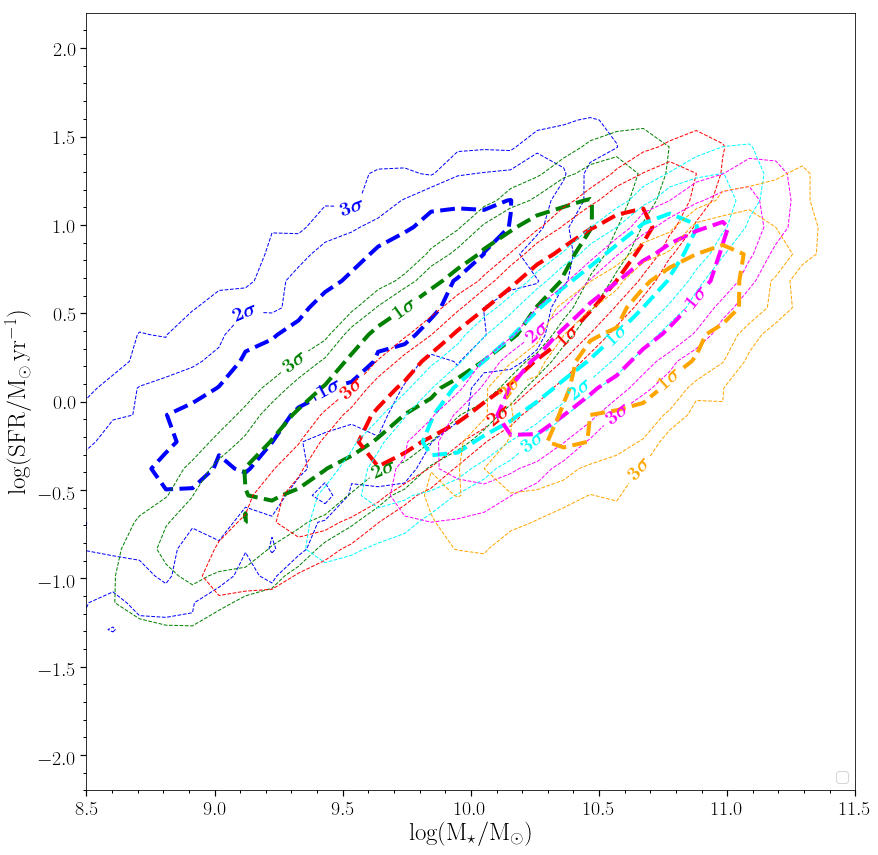}
\caption{Relation between SFR and M$_\star$ for star-forming galaxies colour coded according to the D$\rm _n$(4000) parameter (left panel) and its corresponding confidence limits from 1$\sigma$ to 3$\sigma$ (right panel). All the points and contours of D$\rm _n$(4000) parameter in the figures have the same colours as Fig.~\ref{fig:3_OHNOSFR_d4000}.}
\label{fig:5_SFRM_d4000}
\end{figure*}

Interestingly, a direct consequence of the above findings has been the selection of a sub-sample of galaxies which populate the youngest part of the two diagrams of Fig.~\ref{fig:3_OHNOSFR_d4000}. This group of (8598) galaxies are ``outliers'' of the sample, and cluster around the iso-line $\rm D_n(4000) < 1.1$, on average in both panels of Fig.~\ref{fig:3_OHNOSFR_d4000}. It is for this group of very young galaxies for which the largest changes in the slopes of the O/H {\sl vs.} SFR relation are observed (similar for N/O {\sl vs.} SFR). The galaxies in this group are mainly but not only dwarf galaxies, they sample a range in galaxy stellar masses, as shown; but always showing the largest SFR measured for its stellar mass. This is easily appreciated in Fig.~\ref{fig:5_SFRM_d4000} where the relation between SFR and M$_\star$ for our sample of SDSS star-forming galaxies is presented; galaxies are colour coded according to the D$\rm _n$(4000) parameter. In this figure it can be seen that, for a fixed SFR nearly the entire range of D$\rm _n$(4000) is covered, where the parameter D$\rm _n$(4000) increases as the stellar mass increases, i.e. more massive galaxies have higher D$\rm _n$(4000) values than less massive ones. Similarly, for each fixed stellar mass, a range in SFR is observed where the largest SFR corresponds to the youngest objects. 

In order to explore the nature of these 8598 galaxies we have performed a sanity check in Appendix \ref{sec:sm} to confirm their metallicity and SFR, (see Figures~\ref{fig:A2_sSFRSFRM_d4000}, \ref{fig:A3_spectra}, and \ref{fig:A4_image}) and to gain more insight into the nature of this sub-sample of galaxies. We have checked that the abundances and SFR derived for these galaxies were not affected by aperture corrections, performing a particular study of this subsample. 

% Figure 7__________________________________________________________________
%__________________________________________________________________

\begin{figure*}
\centering
%\begin{minipage}{.48\linewidth}
    \includegraphics[width=.48\textwidth]{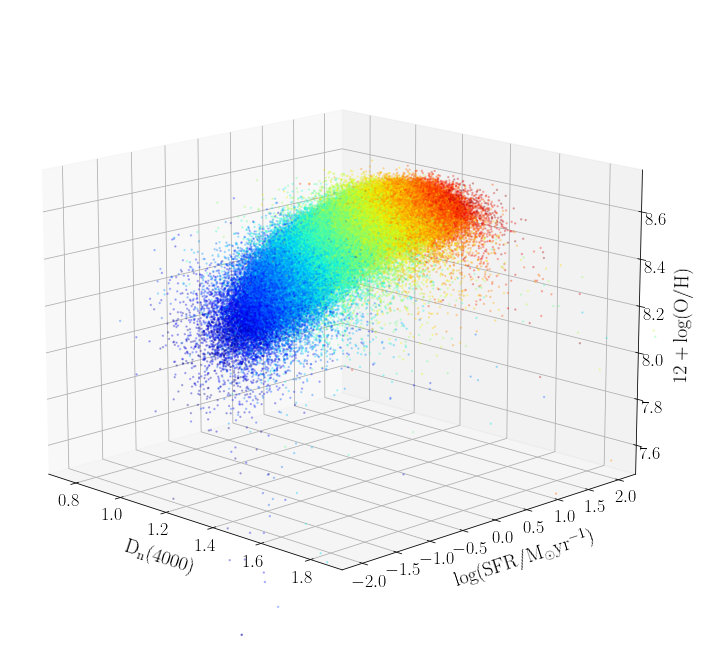}
%\end{minipage}
\hfill
%\begin{minipage}{.48\linewidth}
    \includegraphics[width=.48\textwidth]{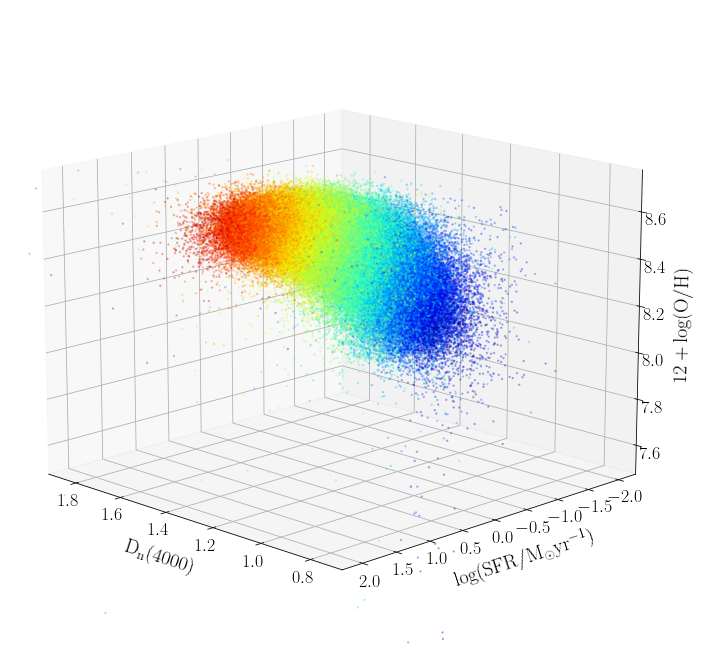}\\
\vfill
    \includegraphics[width=.7\textwidth]{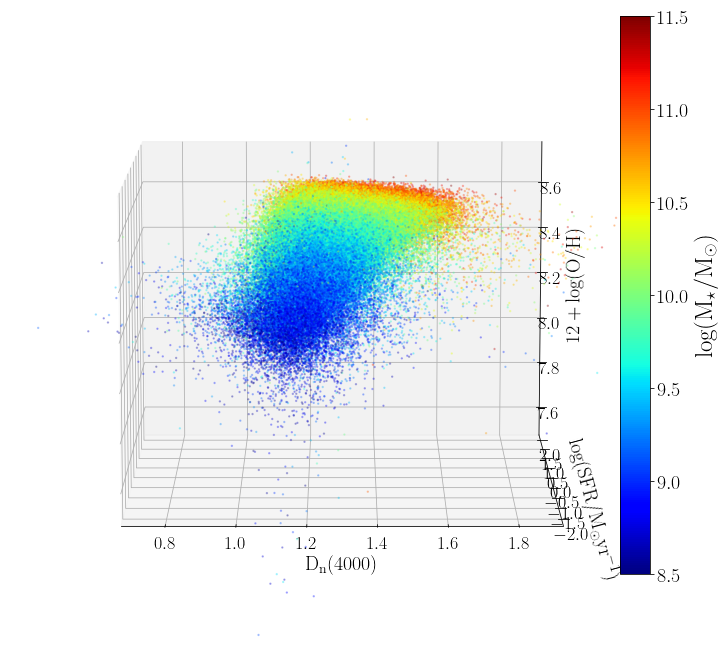}\\
%\end{minipage}
\caption{Population box (D$\rm_{n}$(4000) -- SFR -- 12+log(O/H) diagram) for our sample of star-forming galaxies colour coded by M$_{\star}$.}
\label{fig:6_pop_box}
\end{figure*}

In Fig.~\ref{fig:6_pop_box} we summarise all the results concerning to O/H, SFR and age of the stellar populations, making use of the ``population box" D$\rm_{n}$(4000) -- SFR -- 12+log(O/H) three dimensional diagram, where values found for the star-forming galaxies are colour-coded according to the M$_{\star}$ of each galaxy as shown. A clear correlation between D$\rm_{n}$(4000), SFR and 12+log(O/H) has been found. Galaxies with D$\rm _{n}$(4000) values $ < $ 1.2 span the range of stellar masses $\rm \log{(M_{\star}/M_{\odot}})$ = 8.5 to 11.5, where the majority of the sample galaxies have stellar masses less than $\sim$ 10.5. The less massive galaxies ($\rm \log{(M_{\star}/M_{\odot})} \lesssim 9$) mostly have values of D$\rm _{n}$(4000) less than 1.2, and, as expected, have the lowest 12+log(O/H). On the other hand, the most massive galaxies ($\rm \log{(M_{\star}/M_{\odot})} \gtrsim$ 11) are those with the highest 12+log(O/H) values (solar metallicity) and have D$\rm _{n}$(4000) values higher than 1.1, although most galaxies have values above 1.2. According to the SFR--M$_{\star}$ relation, the higher the stellar mass, the higher the SFR.

% Figure 8__________________________________________________________________
%__________________________________________________________________

The apparent metallicity -- SFR anticorrelation appears to be controversial in the literature; this relation has been interpreted as the consequence of a possible selection effect, and/or possibly resulting from the existence of massive infall of mainly metal poor gas on these galaxies, which would produce a strong enhancement of the SFR and, as a consequence, should dilute the metallicity of the interstellar medium (ISM) \citep{2008ApJ...672L.107E,2010MNRAS.408.2115M,2012MNRAS.422..215Y} giving rise to an anticorrelation between $\rm 12+log(O/H)$ and SFR. This picture seems to be supported by some theoretical work \citep[e.g.][]{2011MNRAS.410.2203S,2013MNRAS.430.2891D}. Though recent observational work has questioned this scenario. We have shown in this work a deep and complex relation between the MZR and SFR which appears to be tuned to the mean age of the galaxy stellar population. The anticorrelation found between oxygen abundance and SFR appears to be significant only for those galaxies hosting the youngest stellar populations. The massive infall scenario, could be invoked to explain these high SFR objects; however for these galaxies we can see that $\rm log(N/O)$ -- SFR also anticorrelate, suggesting a framework more complex than the standard infall scenario where $\rm N/O$ should not be much affected. Further deep spectroscopic observations of the extreme SFR sub-sample could provide valuable information to in order to understand the evolution of this very active star-forming galaxies.

%__________________________________________________________________
%% 4. Summary and conclusions %%%%%%%%%%%%%%%%%%%%%%
\section{Summary and conclusions}
\label{sec:4_conclusions}

In this work we have derived O/H and N/O corrected for aperture effects. These empirical aperture corrections are based on the sample of 165 spiral galaxies from the CALIFA project \citet{2016ApJ...826...71I}. We studied the O/H -- SFR and N/O -- SFR relations, as well as their relation with the parameter D$_n$(4000), a proxy of the age of the stellar populations of the galaxies. We compared the location in the SFR -- M$_\star$ and MZSFR diagrams of the star-forming galaxies according to its value of the D$_n$(4000) for each galaxy.

Our main conclusions are the following:

\begin{enumerate}[i)]

\item A robust stellar mass -- metallicity relation (MZR) locus has been derived consistent with the ``anchoring" points of a selected set of well studied nearby galaxies with a direct derivation of abundance. A complex relation between MZR and SFR across the whole range of galaxy mass and metallicity has been observed, showing a pattern of slope changes of the MZR -- SFR relation in the O/H versus SFR plane, which appears tuned to the age of the stellar population of the galaxies.

\item From the study of the relation between the MZR and SFR, a new dependence between O/H -- SFR has been found with the age of the galaxies' stellar populations. The dependence between the MZR and SFR is strong mainly for star-forming galaxies with high SFR and low $\rm D_n(4000)$. Galaxies with older stellar populations ($\rm D_n(4000) \geq 1.2$) and massive (log(M$_\star$/M$_\odot$) $\gtrsim$ 10.5) tend to saturate, and seem to flatten in the O/H -- SFR relation when 12+log(O/H) $\geq$ 8.55, irrespective of their SFR; which indicates that the metallicity remains constant only depending on the stellar mass of these massive galaxies. We also observe a dependence between N/O -- SFR with the age of the galaxies' stellar populations, where those galaxies with oldest stellar populations and massive seem to flatten when log(N/O) $\geq$ -0.9.

\item On the contrary, for galaxies with younger stellar populations ($\rm D_n(4000) < 1.2$), a negative dependence has been found between O/H and SFR (also for N/O and SFR) for less massive galaxies (log(M$_\star$/M$_\odot$) $\lesssim$ 10.5). This anticorrelation has been interpreted as the existence of massive infall of mainly metal poor gas on the galaxies, which would produce a strong enhancement of the SFR and, as a consequence, should dilute the metallicity of the interstellar medium (ISM).

\item This negative dependence found between O/H and SFR for galaxies with younger stellar populations holds in a metallicity range from typical dwarf values to solar values, covering a wide range of stellar masses, suggesting that galaxies with young stellar populations are not only composed of dwarf galaxies. These young galaxies of our SDSS sample always show the highest SFR measured for their stellar mass.

\item Galaxies with young and old stellar populations are located in different domains in the three dimensional MZSFR diagram: i) galaxies with younger stellar populations ($\rm D_{n}(4000) < 1.2$) are located in the zone where SFR is higher, at fixed 12+log(O/H); and ii) galaxies with older stellar populations ($\rm D_{n}(4000) \ge 1.2$) fall in the area where SFR is lower, at fixed 12+log(O/H).

\end{enumerate}

We have found a population of outliers galaxies in the O/H -- SFR relation (and also in the N/O -- SFR relation), with young stellar population ages. Further characterisation of the physical and chemical properties of these galaxies is necessary to understand how they evolve chemically and to unravel the O/H and N/O relation for these galaxies.

\begin{acknowledgements}
We thank the anonymous referee for very constructive suggestions that have helped us to improve this manuscript. SDP is grateful to the Fonds de Recherche du Québec - Nature et Technologies. SDP, JVM, JIP, CK, and EPM acknowledge financial support from the Spanish Ministerio de Econom\'ia y Competitividad under grants AYA2016-79724-C4-4-P and PID2019-107408GB-C44, from Junta de Andaluc\'ia Excellence Project P18-FR-2664, and also acknowledge support from the State Agency for Research of the Spanish MCIU through the `Center of Excellence Severo Ochoa' award for the Instituto de Astrof\'isica de Andaluc\'ia (SEV-2017-0709). L.S.P acknowledges support within the framework of the program of the NAS of Ukraine Support for the development of priority fields of scientific research (CPCEL 6541230). I.A.Z acknowledges support by the National Academy of Sciences of Ukraine under the Research Laboratory Grant for young scientists No. 0120U100148.\\

Funding for SDSS-III has been provided by the Alfred P. Sloan Foundation, the Participating Institutions, the National Science Foundation, and the U.S. Department of Energy Office of Science. The SDSS-III web site is {\tt \href{http://www.sdss3.org/}{http://www.sdss3.org}}. 

This study makes use of the results based on the Calar Alto Legacy Integral Field Area (CALIFA) survey (\url{http://califa.caha.es/}). 

This research made use of Python ({\tt \href{http://www.python.org}{http://www.python.org}}) and IPython \citep{PER-GRA:2007}; APLpy \citep{2012ascl.soft08017R}; Numpy \citep{2011arXiv1102.1523V}; Pandas \citep{mckinneyprocscipy2010}; of Matplotlib \citep{Hunter:2007}, a suite of open-source Python modules that provides a framework for creating scientific plots. This research made use of Astropy, a community-developed core Python package for Astronomy \citep{2013A&A...558A..33A}. The Astropy web site is {\tt \href{http://www.astropy.org/}{http://www.astropy.org}}.\\
\end{acknowledgements}

\bibliography{ohreferences.bib}

\begin{thebibliography}{101}
\expandafter\ifx\csname natexlab\endcsname\relax\def\natexlab#1{#1}\fi

\bibitem[{{Alam} {et~al.}(2015){Alam}, {Albareti}, {Allende Prieto}, {Anders},
  {Anderson}, {Anderton}, {Andrews}, {Armengaud}, {Aubourg}, {Bailey}, \&
  et~al.}]{2015ApJS..219...12A}
{Alam}, S., {Albareti}, F.~D., {Allende Prieto}, C., {et~al.} 2015, \apjs, 219,
  12

\bibitem[{{Andrews} \& {Martini}(2013)}]{2013ApJ...765..140A}
{Andrews}, B.~H. \& {Martini}, P. 2013, \apj, 765, 140

\bibitem[{{Arellano-C{\'o}rdova} {et~al.}(2021){Arellano-C{\'o}rdova},
  {Esteban}, {Garc{\'\i}a-Rojas}, \&
  {M{\'e}ndez-Delgado}}]{2021MNRAS.502..225A}
{Arellano-C{\'o}rdova}, K.~Z., {Esteban}, C., {Garc{\'\i}a-Rojas}, J., \&
  {M{\'e}ndez-Delgado}, J.~E. 2021, \mnras, 502, 225

\bibitem[{{Asplund} {et~al.}(2009){Asplund}, {Grevesse}, {Sauval}, \&
  {Scott}}]{2009ARA&A..47..481A}
{Asplund}, M., {Grevesse}, N., {Sauval}, A.~J., \& {Scott}, P. 2009, \araa, 47,
  481

\bibitem[{{Astropy Collaboration} {et~al.}(2013){Astropy Collaboration},
  {Robitaille}, {Tollerud}, {Greenfield}, {Droettboom}, {Bray}, {Aldcroft},
  {Davis}, {Ginsburg}, {Price-Whelan}, {Kerzendorf}, {Conley}, {Crighton},
  {Barbary}, {Muna}, {Ferguson}, {Grollier}, {Parikh}, {Nair}, {Unther},
  {Deil}, {Woillez}, {Conseil}, {Kramer}, {Turner}, {Singer}, {Fox}, {Weaver},
  {Zabalza}, {Edwards}, {Azalee Bostroem}, {Burke}, {Casey}, {Crawford},
  {Dencheva}, {Ely}, {Jenness}, {Labrie}, {Lim}, {Pierfederici}, {Pontzen},
  {Ptak}, {Refsdal}, {Servillat}, \& {Streicher}}]{2013A&A...558A..33A}
{Astropy Collaboration}, {Robitaille}, T.~P., {Tollerud}, E.~J., {et~al.} 2013,
  \aap, 558, A33

\bibitem[{{Balogh} {et~al.}(1999){Balogh}, {Morris}, {Yee}, {Carlberg}, \&
  {Ellingson}}]{1999ApJ...527...54B}
{Balogh}, M.~L., {Morris}, S.~L., {Yee}, H.~K.~C., {Carlberg}, R.~G., \&
  {Ellingson}, E. 1999, \apj, 527, 54

\bibitem[{{Barrera-Ballesteros} {et~al.}(2018){Barrera-Ballesteros}, {Heckman},
  {S{\'a}nchez}, {Zakamska}, {Cleary}, {Zhu}, {Brinkmann}, {Drory}, \& {THE
  MaNGA TEAM}}]{2018ApJ...852...74B}
{Barrera-Ballesteros}, J.~K., {Heckman}, T., {S{\'a}nchez}, S.~F., {et~al.}
  2018, \apj, 852, 74

\bibitem[{{Barrera-Ballesteros} {et~al.}(2017){Barrera-Ballesteros},
  {S{\'a}nchez}, {Heckman}, {Blanc}, \& {The MaNGA Team}}]{2017ApJ...844...80B}
{Barrera-Ballesteros}, J.~K., {S{\'a}nchez}, S.~F., {Heckman}, T., {Blanc},
  G.~A., \& {The MaNGA Team}. 2017, \apj, 844, 80

\bibitem[{{Barrera-Ballesteros} {et~al.}(2020){Barrera-Ballesteros}, {Utomo},
  {Bolatto}, {S{\'a}nchez}, {Vogel}, {Wong}, {Levy}, {Colombo}, {Kalinova},
  {Teuben}, {Garc{\'\i}a-Benito}, {Husemann}, {Mast}, \&
  {Blitz}}]{2020MNRAS.492.2651B}
{Barrera-Ballesteros}, J.~K., {Utomo}, D., {Bolatto}, A.~D., {et~al.} 2020,
  \mnras, 492, 2651

\bibitem[{{Belfiore} {et~al.}(2016){Belfiore}, {Maiolino}, {Maraston},
  {Emsellem}, {Bershady}, {Masters}, {Yan}, {Bizyaev}, {Boquien}, {Brownstein},
  {Bundy}, {Drory}, {Heckman}, {Law}, {Roman-Lopes}, {Pan}, {Stanghellini},
  {Thomas}, {Weijmans}, \& {Westfall}}]{2016MNRAS.461.3111B}
{Belfiore}, F., {Maiolino}, R., {Maraston}, C., {et~al.} 2016, \mnras, 461,
  3111

\bibitem[{{Berg} {et~al.}(2015){Berg}, {Skillman}, {Croxall}, {Pogge},
  {Moustakas}, \& {Johnson-Groh}}]{2015ApJ...806...16B}
{Berg}, D.~A., {Skillman}, E.~D., {Croxall}, K.~V., {et~al.} 2015, \apj, 806,
  16

\bibitem[{{Bresolin} {et~al.}(2016){Bresolin}, {Kudritzki}, {Urbaneja},
  {Gieren}, {Ho}, \& {Pietrzy{\'n}ski}}]{2016ApJ...830...64B}
{Bresolin}, F., {Kudritzki}, R.-P., {Urbaneja}, M.~A., {et~al.} 2016, \apj,
  830, 64

\bibitem[{{Brinchmann} {et~al.}(2004){Brinchmann}, {Charlot}, {White},
  {Tremonti}, {Kauffmann}, {Heckman}, \& {Brinkmann}}]{2004MNRAS.351.1151B}
{Brinchmann}, J., {Charlot}, S., {White}, S.~D.~M., {et~al.} 2004, \mnras, 351,
  1151

\bibitem[{{Brown} {et~al.}(2016){Brown}, {Martini}, \&
  {Andrews}}]{2016MNRAS.458.1529B}
{Brown}, J.~S., {Martini}, P., \& {Andrews}, B.~H. 2016, \mnras, 458, 1529

\bibitem[{{Bundy} {et~al.}(2015){Bundy}, {Bershady}, {Law}, {Yan}, {Drory},
  {MacDonald}, {Wake}, {Cherinka}, {S{\'a}nchez-Gallego}, {Weijmans}, {Thomas},
  {Tremonti}, {Masters}, {Coccato}, {Diamond-Stanic}, {Arag{\'o}n-Salamanca},
  {Avila-Reese}, {Badenes}, {Falc{\'o}n-Barroso}, {Belfiore}, {Bizyaev},
  {Blanc}, {Bland-Hawthorn}, {Blanton}, {Brownstein}, {Byler}, {Cappellari},
  {Conroy}, {Dutton}, {Emsellem}, {Etherington}, {Frinchaboy}, {Fu}, {Gunn},
  {Harding}, {Johnston}, {Kauffmann}, {Kinemuchi}, {Klaene}, {Knapen},
  {Leauthaud}, {Li}, {Lin}, {Maiolino}, {Malanushenko}, {Malanushenko}, {Mao},
  {Maraston}, {McDermid}, {Merrifield}, {Nichol}, {Oravetz}, {Pan}, {Parejko},
  {Sanchez}, {Schlegel}, {Simmons}, {Steele}, {Steinmetz}, {Thanjavur},
  {Thompson}, {Tinker}, {van den Bosch}, {Westfall}, {Wilkinson}, {Wright},
  {Xiao}, \& {Zhang}}]{2015ApJ...798....7B}
{Bundy}, K., {Bershady}, M.~A., {Law}, D.~R., {et~al.} 2015, \apj, 798, 7

\bibitem[{{Chabrier}(2003)}]{2003ApJ...586L.133C}
{Chabrier}, G. 2003, \apjl, 586, L133

\bibitem[{{Cook} {et~al.}(2014){Cook}, {Dale}, {Johnson}, {Van Zee}, {Lee},
  {Kennicutt}, {Calzetti}, {Staudaher}, \& {Engelbracht}}]{2014MNRAS.445..899C}
{Cook}, D.~O., {Dale}, D.~A., {Johnson}, B.~D., {et~al.} 2014, \mnras, 445, 899

\bibitem[{{Cresci} {et~al.}(2019){Cresci}, {Mannucci}, \&
  {Curti}}]{2019A&A...627A..42C}
{Cresci}, G., {Mannucci}, F., \& {Curti}, M. 2019, \aap, 627, A42

\bibitem[{{Cresci} {et~al.}(2012){Cresci}, {Mannucci}, {Sommariva}, {Maiolino},
  {Marconi}, \& {Brusa}}]{2012MNRAS.421..262C}
{Cresci}, G., {Mannucci}, F., {Sommariva}, V., {et~al.} 2012, \mnras, 421, 262

\bibitem[{{Croxall} {et~al.}(2016){Croxall}, {Pogge}, {Berg}, {Skillman}, \&
  {Moustakas}}]{2016ApJ...830....4C}
{Croxall}, K.~V., {Pogge}, R.~W., {Berg}, D.~A., {Skillman}, E.~D., \&
  {Moustakas}, J. 2016, \apj, 830, 4

\bibitem[{{Curti} {et~al.}(2017){Curti}, {Cresci}, {Mannucci}, {Marconi},
  {Maiolino}, \& {Esposito}}]{2017MNRAS.465.1384C}
{Curti}, M., {Cresci}, G., {Mannucci}, F., {et~al.} 2017, \mnras, 465, 1384

\bibitem[{{Curti} {et~al.}(2020){Curti}, {Mannucci}, {Cresci}, \&
  {Maiolino}}]{2020MNRAS.491..944C}
{Curti}, M., {Mannucci}, F., {Cresci}, G., \& {Maiolino}, R. 2020, \mnras, 491,
  944

\bibitem[{{Dav{\'e}} {et~al.}(2011){Dav{\'e}}, {Finlator}, \&
  {Oppenheimer}}]{2011MNRAS.416.1354D}
{Dav{\'e}}, R., {Finlator}, K., \& {Oppenheimer}, B.~D. 2011, \mnras, 416, 1354

\bibitem[{{Dayal} {et~al.}(2013){Dayal}, {Ferrara}, \&
  {Dunlop}}]{2013MNRAS.430.2891D}
{Dayal}, P., {Ferrara}, A., \& {Dunlop}, J.~S. 2013, \mnras, 430, 2891

\bibitem[{{De Rossi} {et~al.}(2017){De Rossi}, {Bower}, {Font}, {Schaye}, \&
  {Theuns}}]{2017MNRAS.472.3354D}
{De Rossi}, M.~E., {Bower}, R.~G., {Font}, A.~S., {Schaye}, J., \& {Theuns}, T.
  2017, \mnras, 472, 3354

\bibitem[{{Duarte Puertas} {et~al.}(2017){Duarte Puertas}, {Vilchez},
  {Iglesias-P{\'a}ramo}, {Kehrig}, {P{\'e}rez-Montero}, \&
  {Rosales-Ortega}}]{2017A&A...599A..71D}
{Duarte Puertas}, S., {Vilchez}, J.~M., {Iglesias-P{\'a}ramo}, J., {et~al.}
  2017, \aap, 599, A71

\bibitem[{{Edmunds} \& {Pagel}(1978)}]{1978MNRAS.185P..77E}
{Edmunds}, M.~G. \& {Pagel}, B.~E.~J. 1978, \mnras, 185, 77P

\bibitem[{{Ellison} {et~al.}(2008){Ellison}, {Patton}, {Simard}, \&
  {McConnachie}}]{2008ApJ...672L.107E}
{Ellison}, S.~L., {Patton}, D.~R., {Simard}, L., \& {McConnachie}, A.~W. 2008,
  \apjl, 672, L107

\bibitem[{{Ellison} {et~al.}(2009){Ellison}, {Simard}, {Cowan}, {Baldry},
  {Patton}, \& {McConnachie}}]{2009MNRAS.396.1257E}
{Ellison}, S.~L., {Simard}, L., {Cowan}, N.~B., {et~al.} 2009, \mnras, 396,
  1257

\bibitem[{{Fern{\'a}ndez-Mart{\'\i}n}
  {et~al.}(2017){Fern{\'a}ndez-Mart{\'\i}n}, {P{\'e}rez-Montero},
  {V{\'\i}lchez}, \& {Mampaso}}]{2017A&A...597A..84F}
{Fern{\'a}ndez-Mart{\'\i}n}, A., {P{\'e}rez-Montero}, E., {V{\'\i}lchez},
  J.~M., \& {Mampaso}, A. 2017, \aap, 597, A84

\bibitem[{{Finlator} \& {Dav{\'e}}(2008)}]{2008MNRAS.385.2181F}
{Finlator}, K. \& {Dav{\'e}}, R. 2008, \mnras, 385, 2181

\bibitem[{{Fisher} \& {Drory}(2011)}]{2011ApJ...733L..47F}
{Fisher}, D.~B. \& {Drory}, N. 2011, \apjl, 733, L47

\bibitem[{{Gallazzi} {et~al.}(2005){Gallazzi}, {Charlot}, {Brinchmann},
  {White}, \& {Tremonti}}]{2005MNRAS.362...41G}
{Gallazzi}, A., {Charlot}, S., {Brinchmann}, J., {White}, S.~D.~M., \&
  {Tremonti}, C.~A. 2005, \mnras, 362, 41

\bibitem[{{Garc{\'{\i}}a-Benito} {et~al.}(2015){Garc{\'{\i}}a-Benito},
  {Zibetti}, {S{\'a}nchez}, {Husemann}, {de Amorim}, {Castillo-Morales}, {Cid
  Fernandes}, {Ellis}, {Falc{\'o}n-Barroso}, {Galbany}, {Gil de Paz},
  {Gonz{\'a}lez Delgado}, {Lacerda}, {L{\'o}pez-Fernandez}, {de
  Lorenzo-C{\'a}ceres}, {Lyubenova}, {Marino}, {Mast}, {Mendoza}, {P{\'e}rez},
  {Vale Asari}, {Aguerri}, {Ascasibar}, {Bekerait*error*{\.e}},
  {Bland-Hawthorn}, {Barrera-Ballesteros}, {Bomans}, {Cano-D{\'{\i}}az},
  {Catal{\'a}n-Torrecilla}, {Cortijo}, {Delgado-Inglada}, {Demleitner},
  {Dettmar}, {D{\'{\i}}az}, {Florido}, {Gallazzi}, {Garc{\'{\i}}a-Lorenzo},
  {Gomes}, {Holmes}, {Iglesias-P{\'a}ramo}, {Jahnke}, {Kalinova}, {Kehrig},
  {Kennicutt}, {L{\'o}pez-S{\'a}nchez}, {M{\'a}rquez}, {Masegosa}, {Meidt},
  {Mendez-Abreu}, {Moll{\'a}}, {Monreal-Ibero}, {Morisset}, {del Olmo},
  {Papaderos}, {P{\'e}rez}, {Quirrenbach}, {Rosales-Ortega}, {Roth},
  {Ruiz-Lara}, {S{\'a}nchez-Bl{\'a}zquez}, {S{\'a}nchez-Menguiano}, {Singh},
  {Spekkens}, {Stanishev}, {Torres-Papaqui}, {van de Ven}, {Vilchez},
  {Walcher}, {Wild}, {Wisotzki}, {Ziegler}, {Alves}, {Barrado}, {Quintana}, \&
  {Aceituno}}]{2015A&A...576A.135G}
{Garc{\'{\i}}a-Benito}, R., {Zibetti}, S., {S{\'a}nchez}, S.~F., {et~al.} 2015,
  \aap, 576, A135

\bibitem[{{Garnett}(2002)}]{2002ApJ...581.1019G}
{Garnett}, D.~R. 2002, \apj, 581, 1019

\bibitem[{{Gomes} {et~al.}(2016){Gomes}, {Papaderos}, {V{\'{\i}}lchez},
  {Kehrig}, {Iglesias-P{\'a}ramo}, {Breda}, {Lehnert}, {S{\'a}nchez},
  {Ziegler}, {Dos Reis}, {Bland-Hawthorn}, {Galbany}, {Bomans},
  {Rosales-Ortega}, {Walcher}, {Garc{\'{\i}}a-Benito}, {M{\'a}rquez}, {Del
  Olmo}, {Moll{\'a}}, {Marino}, {Catal{\'a}n-Torrecilla}, {Gonz{\'a}lez
  Delgado}, {L{\'o}pez-S{\'a}nchez}, \& {Califa
  Collaboration}}]{2016A&A...586A..22G}
{Gomes}, J.~M., {Papaderos}, P., {V{\'{\i}}lchez}, J.~M., {et~al.} 2016, \aap,
  586, A22

\bibitem[{{Hernandez} {et~al.}(2019){Hernandez}, {Larsen}, {Aloisi}, {Berg},
  {Blair}, {Fox}, {Heckman}, {James}, {Long}, {Skillman}, \&
  {Whitmore}}]{2019ApJ...872..116H}
{Hernandez}, S., {Larsen}, S., {Aloisi}, A., {et~al.} 2019, \apj, 872, 116

\bibitem[{{Hughes} {et~al.}(2013){Hughes}, {Cortese}, {Boselli}, {Gavazzi}, \&
  {Davies}}]{2013A&A...550A.115H}
{Hughes}, T.~M., {Cortese}, L., {Boselli}, A., {Gavazzi}, G., \& {Davies},
  J.~I. 2013, \aap, 550, A115

\bibitem[{{Hunt} {et~al.}(2016){Hunt}, {Dayal}, {Magrini}, \&
  {Ferrara}}]{2016MNRAS.463.2002H}
{Hunt}, L., {Dayal}, P., {Magrini}, L., \& {Ferrara}, A. 2016, \mnras, 463,
  2002

\bibitem[{Hunter(2007)}]{Hunter:2007}
Hunter, J.~D. 2007, Computing In Science \& Engineering, 9, 90

\bibitem[{{Iglesias-P{\'a}ramo} {et~al.}(2013){Iglesias-P{\'a}ramo},
  {V{\'{\i}}lchez}, {Galbany}, {S{\'a}nchez}, {Rosales-Ortega}, {Mast},
  {Garc{\'{\i}}a-Benito}, {Husemann}, {Aguerri}, {Alves}, {Bekerait{\'e}},
  {Bland-Hawthorn}, {Catal{\'a}n-Torrecilla}, {de Amorim}, {de
  Lorenzo-C{\'a}ceres}, {Ellis}, {Falc{\'o}n-Barroso}, {Flores}, {Florido},
  {Gallazzi}, {Gomes}, {Gonz{\'a}lez Delgado}, {Haines},
  {Hern{\'a}ndez-Fern{\'a}ndez}, {Kehrig}, {L{\'o}pez-S{\'a}nchez},
  {Lyubenova}, {Marino}, {Moll{\'a}}, {Monreal-Ibero}, {Mour{\~a}o},
  {Papaderos}, {Rodrigues}, {S{\'a}nchez-Bl{\'a}zquez}, {Spekkens},
  {Stanishev}, {van de Ven}, {Walcher}, {Wisotzki}, {Zibetti}, \&
  {Ziegler}}]{2013A&A...553L...7I}
{Iglesias-P{\'a}ramo}, J., {V{\'{\i}}lchez}, J.~M., {Galbany}, L., {et~al.}
  2013, \aap, 553, L7

\bibitem[{{Iglesias-P{\'a}ramo} {et~al.}(2016){Iglesias-P{\'a}ramo},
  {V{\'{\i}}lchez}, {Rosales-Ortega}, {S{\'a}nchez}, {Duarte Puertas},
  {Petropoulou}, {Gil de Paz}, {Galbany}, {Moll{\'a}},
  {Catal{\'a}n-Torrecilla}, {Castillo Morales}, {Mast}, {Husemann},
  {Garc{\'{\i}}a-Benito}, {Mendoza}, {Kehrig}, {P{\'e}rez-Montero},
  {Papaderos}, {Gomes}, {Walcher}, {Gonz{\'a}lez Delgado}, {Marino},
  {L{\'o}pez-S{\'a}nchez}, {Ziegler}, {Flores}, \&
  {Alves}}]{2016ApJ...826...71I}
{Iglesias-P{\'a}ramo}, J., {V{\'{\i}}lchez}, J.~M., {Rosales-Ortega}, F.~F.,
  {et~al.} 2016, \apj, 826, 71

\bibitem[{{Kang} {et~al.}(2016){Kang}, {Zhang}, {Chang}, {Wang}, \&
  {Cheng}}]{2016A&A...585A..20K}
{Kang}, X., {Zhang}, F., {Chang}, R., {Wang}, L., \& {Cheng}, L. 2016, \aap,
  585, A20

\bibitem[{{Kashino} {et~al.}(2016){Kashino}, {Renzini}, {Silverman}, \&
  {Daddi}}]{2016ApJ...823L..24K}
{Kashino}, D., {Renzini}, A., {Silverman}, J.~D., \& {Daddi}, E. 2016, \apjl,
  823, L24

\bibitem[{{Kauffmann} {et~al.}(2003){Kauffmann}, {Heckman}, {White}, {Charlot},
  {Tremonti}, {Brinchmann}, {Bruzual}, {Peng}, {Seibert}, {Bernardi},
  {Blanton}, {Brinkmann}, {Castander}, {Cs{\'a}bai}, {Fukugita}, {Ivezic},
  {Munn}, {Nichol}, {Padmanabhan}, {Thakar}, {Weinberg}, \&
  {York}}]{2003MNRAS.341...33K}
{Kauffmann}, G., {Heckman}, T.~M., {White}, S.~D.~M., {et~al.} 2003, \mnras,
  341, 33

\bibitem[{{Kojima} {et~al.}(2017){Kojima}, {Ouchi}, {Nakajima}, {Shibuya},
  {Harikane}, \& {Ono}}]{2017PASJ...69...44K}
{Kojima}, T., {Ouchi}, M., {Nakajima}, K., {et~al.} 2017, \pasj, 69, 44

\bibitem[{{Kroupa}(2001)}]{2001MNRAS.322..231K}
{Kroupa}, P. 2001, \mnras, 322, 231

\bibitem[{{Kudritzki} {et~al.}(2012){Kudritzki}, {Urbaneja}, {Gazak},
  {Bresolin}, {Przybilla}, {Gieren}, \&
  {Pietrzy{\'n}ski}}]{2012ApJ...747...15K}
{Kudritzki}, R.-P., {Urbaneja}, M.~A., {Gazak}, Z., {et~al.} 2012, \apj, 747,
  15

\bibitem[{{Lara-L{\'o}pez} {et~al.}(2010{\natexlab{a}}){Lara-L{\'o}pez},
  {Bongiovanni}, {Cepa}, {P{\'e}rez Garc{\'{\i}}a}, {S{\'a}nchez-Portal},
  {Casta{\~n}eda}, {Fern{\'a}ndez Lorenzo}, \&
  {Povi{\'c}}}]{2010A&A...519A..31L}
{Lara-L{\'o}pez}, M.~A., {Bongiovanni}, A., {Cepa}, J., {et~al.}
  2010{\natexlab{a}}, \aap, 519, A31

\bibitem[{{Lara-L{\'o}pez} {et~al.}(2010{\natexlab{b}}){Lara-L{\'o}pez},
  {Cepa}, {Bongiovanni}, {P{\'e}rez Garc{\'{\i}}a}, {Ederoclite},
  {Casta{\~n}eda}, {Fern{\'a}ndez Lorenzo}, {Povi{\'c}}, \&
  {S{\'a}nchez-Portal}}]{2010A&A...521L..53L}
{Lara-L{\'o}pez}, M.~A., {Cepa}, J., {Bongiovanni}, A., {et~al.}
  2010{\natexlab{b}}, \aap, 521, L53

\bibitem[{{Lee} {et~al.}(2006){Lee}, {Skillman}, {Cannon}, {Jackson}, {Gehrz},
  {Polomski}, \& {Woodward}}]{2006ApJ...647..970L}
{Lee}, H., {Skillman}, E.~D., {Cannon}, J.~M., {et~al.} 2006, \apj, 647, 970

\bibitem[{{Lequeux} {et~al.}(1979){Lequeux}, {Peimbert}, {Rayo}, {Serrano}, \&
  {Torres-Peimbert}}]{1979A&A....80..155L}
{Lequeux}, J., {Peimbert}, M., {Rayo}, J.~F., {Serrano}, A., \&
  {Torres-Peimbert}, S. 1979, \aap, 80, 155

\bibitem[{{Licquia} \& {Newman}(2015)}]{2015ApJ...806...96L}
{Licquia}, T.~C. \& {Newman}, J.~A. 2015, \apj, 806, 96

\bibitem[{{Magrini} {et~al.}(2017){Magrini}, {Gon{\c c}alves}, \&
  {Vajgel}}]{2017MNRAS.464..739M}
{Magrini}, L., {Gon{\c c}alves}, D.~R., \& {Vajgel}, B. 2017, \mnras, 464, 739

\bibitem[{{Maiolino} \& {Mannucci}(2019)}]{2019A&ARv..27....3M}
{Maiolino}, R. \& {Mannucci}, F. 2019, \aapr, 27, 3

\bibitem[{{Mannucci} {et~al.}(2010){Mannucci}, {Cresci}, {Maiolino}, {Marconi},
  \& {Gnerucci}}]{2010MNRAS.408.2115M}
{Mannucci}, F., {Cresci}, G., {Maiolino}, R., {Marconi}, A., \& {Gnerucci}, A.
  2010, \mnras, 408, 2115

\bibitem[{{Mast} {et~al.}(2014){Mast}, {Rosales-Ortega}, {S{\'a}nchez},
  {V{\'{\i}}lchez}, {Iglesias-Paramo}, {Walcher}, {Husemann}, {M{\'a}rquez},
  {Marino}, {Kennicutt}, {Monreal-Ibero}, {Galbany}, {de Lorenzo-C{\'a}ceres},
  {Mendez-Abreu}, {Kehrig}, {del Olmo}, {Rela{\~n}o}, {Wisotzki},
  {M{\'a}rmol-Queralt{\'o}}, {Bekerait{\`e}}, {Papaderos}, {Wild}, {Aguerri},
  {Falc{\'o}n-Barroso}, {Bomans}, {Ziegler}, {Garc{\'{\i}}a-Lorenzo},
  {Bland-Hawthorn}, {L{\'o}pez-S{\'a}nchez}, \& {van de
  Ven}}]{2014A&A...561A.129M}
{Mast}, D., {Rosales-Ortega}, F.~F., {S{\'a}nchez}, S.~F., {et~al.} 2014, \aap,
  561, A129

\bibitem[{{Mateus} {et~al.}(2006){Mateus}, {Sodr{\'e}}, {Cid Fernandes},
  {Stasi{\'n}ska}, {Schoenell}, \& {Gomes}}]{2006MNRAS.370..721M}
{Mateus}, A., {Sodr{\'e}}, L., {Cid Fernandes}, R., {et~al.} 2006, \mnras, 370,
  721

\bibitem[{McKinney(2010)}]{mckinneyprocscipy2010}
McKinney, W. 2010, in Proceedings of the 9th Python in Science Conference, ed.
  S.~van~der Walt \& J.~Millman, 51 -- 56

\bibitem[{{Moll{\'a}} {et~al.}(2006){Moll{\'a}}, {V{\'{\i}}lchez},
  {Gavil{\'a}n}, \& {D{\'{\i}}az}}]{2006MNRAS.372.1069M}
{Moll{\'a}}, M., {V{\'{\i}}lchez}, J.~M., {Gavil{\'a}n}, M., \& {D{\'{\i}}az},
  A.~I. 2006, \mnras, 372, 1069

\bibitem[{{M{\o}ller} {et~al.}(2013){M{\o}ller}, {Fynbo}, {Ledoux}, \&
  {Nilsson}}]{2013MNRAS.430.2680M}
{M{\o}ller}, P., {Fynbo}, J.~P.~U., {Ledoux}, C., \& {Nilsson}, K.~K. 2013,
  \mnras, 430, 2680

\bibitem[{{Mouhcine} {et~al.}(2008){Mouhcine}, {Gibson}, {Renda}, \&
  {Kawata}}]{2008A&A...486..711M}
{Mouhcine}, M., {Gibson}, B.~K., {Renda}, A., \& {Kawata}, D. 2008, \aap, 486,
  711

\bibitem[{{Okamoto} {et~al.}(2017){Okamoto}, {Nagashima}, {Lacey}, \&
  {Frenk}}]{2017MNRAS.464.4866O}
{Okamoto}, T., {Nagashima}, M., {Lacey}, C.~G., \& {Frenk}, C.~S. 2017, \mnras,
  464, 4866

\bibitem[{{Peng} \& {Maiolino}(2014)}]{2014MNRAS.438..262P}
{Peng}, Y.-j. \& {Maiolino}, R. 2014, \mnras, 438, 262

\bibitem[{P\'erez \& Granger(2007)}]{PER-GRA:2007}
P\'erez, F. \& Granger, B.~E. 2007, Computing in Science and Engineering, 9, 21

\bibitem[{{P{\'e}rez-Montero}(2014)}]{2014MNRAS.441.2663P}
{P{\'e}rez-Montero}, E. 2014, \mnras, 441, 2663

\bibitem[{{P{\'e}rez-Montero} \& {Contini}(2009)}]{2009MNRAS.398..949P}
{P{\'e}rez-Montero}, E. \& {Contini}, T. 2009, \mnras, 398, 949

\bibitem[{{P{\'e}rez-Montero} {et~al.}(2013){P{\'e}rez-Montero}, {Contini},
  {Lamareille}, {Maier}, {Carollo}, {Kneib}, {Le F{\`e}vre}, {Lilly},
  {Mainieri}, {Renzini}, {Scodeggio}, {Zamorani}, {Bardelli}, {Bolzonella},
  {Bongiorno}, {Caputi}, {Cucciati}, {de la Torre}, {de Ravel}, {Franzetti},
  {Garilli}, {Iovino}, {Kampczyk}, {Knobel}, {Kova{\v c}}, {Le Borgne}, {Le
  Brun}, {Mignoli}, {Pell{\`o}}, {Peng}, {Presotto}, {Ricciardelli},
  {Silverman}, {Tanaka}, {Tasca}, {Tresse}, {Vergani}, \&
  {Zucca}}]{2013A&A...549A..25P}
{P{\'e}rez-Montero}, E., {Contini}, T., {Lamareille}, F., {et~al.} 2013, \aap,
  549, A25

\bibitem[{{Petropoulou} {et~al.}(2012){Petropoulou}, {V{\'{\i}}lchez}, \&
  {Iglesias-P{\'a}ramo}}]{2012ApJ...749..133P}
{Petropoulou}, V., {V{\'{\i}}lchez}, J., \& {Iglesias-P{\'a}ramo}, J. 2012,
  \apj, 749, 133

\bibitem[{{Petropoulou} {et~al.}(2011){Petropoulou}, {V{\'{\i}}lchez},
  {Iglesias-P{\'a}ramo}, {Papaderos}, {Magrini}, {Cedr{\'e}s}, \&
  {Reverte}}]{2011ApJ...734...32P}
{Petropoulou}, V., {V{\'{\i}}lchez}, J., {Iglesias-P{\'a}ramo}, J., {et~al.}
  2011, \apj, 734, 32

\bibitem[{{Pilyugin} \& {Grebel}(2016)}]{2016MNRAS.457.3678P}
{Pilyugin}, L.~S. \& {Grebel}, E.~K. 2016, \mnras, 457, 3678

\bibitem[{{Pilyugin} {et~al.}(2017){Pilyugin}, {Grebel}, {Zinchenko},
  {Nefedyev}, \& {Mattsson}}]{2017MNRAS.465.1358P}
{Pilyugin}, L.~S., {Grebel}, E.~K., {Zinchenko}, I.~A., {Nefedyev}, Y.~A., \&
  {Mattsson}, L. 2017, \mnras, 465, 1358

\bibitem[{{Pilyugin} {et~al.}(2013){Pilyugin}, {Lara-L{\'o}pez}, {Grebel},
  {Kehrig}, {Zinchenko}, {L{\'o}pez-S{\'a}nchez}, {V{\'{\i}}lchez}, \&
  {Mattsson}}]{2013MNRAS.432.1217P}
{Pilyugin}, L.~S., {Lara-L{\'o}pez}, M.~A., {Grebel}, E.~K., {et~al.} 2013,
  \mnras, 432, 1217

\bibitem[{{Pilyugin} {et~al.}(2003){Pilyugin}, {Thuan}, \&
  {V{\'{\i}}lchez}}]{2003A&A...397..487P}
{Pilyugin}, L.~S., {Thuan}, T.~X., \& {V{\'{\i}}lchez}, J.~M. 2003, \aap, 397,
  487

\bibitem[{{Pilyugin} {et~al.}(2007){Pilyugin}, {Thuan}, \&
  {V{\'{\i}}lchez}}]{2007MNRAS.376..353P}
{Pilyugin}, L.~S., {Thuan}, T.~X., \& {V{\'{\i}}lchez}, J.~M. 2007, \mnras,
  376, 353

\bibitem[{{Pilyugin} {et~al.}(2004){Pilyugin}, {V{\'{\i}}lchez}, \&
  {Contini}}]{2004A&A...425..849P}
{Pilyugin}, L.~S., {V{\'{\i}}lchez}, J.~M., \& {Contini}, T. 2004, \aap, 425,
  849

\bibitem[{{Pilyugin} {et~al.}(2012){Pilyugin}, {V{\'{\i}}lchez}, {Mattsson}, \&
  {Thuan}}]{2012MNRAS.421.1624P}
{Pilyugin}, L.~S., {V{\'{\i}}lchez}, J.~M., {Mattsson}, L., \& {Thuan}, T.~X.
  2012, \mnras, 421, 1624

\bibitem[{{Robitaille} \& {Bressert}(2012)}]{2012ascl.soft08017R}
{Robitaille}, T. \& {Bressert}, E. 2012, {APLpy: Astronomical Plotting Library
  in Python}, Astrophysics Source Code Library

\bibitem[{{Romeo Velon{\`a}} {et~al.}(2013){Romeo Velon{\`a}}, {Sommer-Larsen},
  {Napolitano}, {Antonuccio-Delogu}, {Cielo}, {Gavignaud}, \&
  {Meza}}]{2013ApJ...770..155R}
{Romeo Velon{\`a}}, A.~D., {Sommer-Larsen}, J., {Napolitano}, N.~R., {et~al.}
  2013, \apj, 770, 155

\bibitem[{{Sackett}(1997)}]{1997ApJ...483..103S}
{Sackett}, P.~D. 1997, \apj, 483, 103

\bibitem[{{Sakstein} {et~al.}(2011){Sakstein}, {Pipino}, {Devriendt}, \&
  {Maiolino}}]{2011MNRAS.410.2203S}
{Sakstein}, J., {Pipino}, A., {Devriendt}, J.~E.~G., \& {Maiolino}, R. 2011,
  \mnras, 410, 2203

\bibitem[{{Salim} {et~al.}(2014){Salim}, {Lee}, {Ly}, {Brinchmann}, {Dav{\'e}},
  {Dickinson}, {Salzer}, \& {Charlot}}]{2014ApJ...797..126S}
{Salim}, S., {Lee}, J.~C., {Ly}, C., {et~al.} 2014, \apj, 797, 126

\bibitem[{{Salim} {et~al.}(2007){Salim}, {Rich}, {Charlot}, {Brinchmann},
  {Johnson}, {Schiminovich}, {Seibert}, {Mallery}, {Heckman}, {Forster},
  {Friedman}, {Martin}, {Morrissey}, {Neff}, {Small}, {Wyder}, {Bianchi},
  {Donas}, {Lee}, {Madore}, {Milliard}, {Szalay}, {Welsh}, \&
  {Yi}}]{2007ApJS..173..267S}
{Salim}, S., {Rich}, R.~M., {Charlot}, S., {et~al.} 2007, \apjs, 173, 267

\bibitem[{{S{\'a}nchez}(2020)}]{2020ARA&A..5812120S}
{S{\'a}nchez}, S.~F. 2020, \araa, 58, annurev

\bibitem[{{S{\'a}nchez} {et~al.}(2019){S{\'a}nchez}, {Barrera-Ballesteros},
  {L{\'o}pez-Cob{\'a}}, {Brough}, {Bryant}, {Bland -Hawthorn}, {Croom}, {van de
  Sande}, {Cortese}, {Goodwin}, {Lawrence}, {L{\'o}pez-S{\'a}nchez}, {Sweet},
  {Owers}, {Richards}, \& {Walcher}}]{2019MNRAS.484.3042S}
{S{\'a}nchez}, S.~F., {Barrera-Ballesteros}, J.~K., {L{\'o}pez-Cob{\'a}}, C.,
  {et~al.} 2019, \mnras, 484, 3042

\bibitem[{{S{\'a}nchez} {et~al.}(2017){S{\'a}nchez}, {Barrera-Ballesteros},
  {S{\'a}nchez-Menguiano}, {Walcher}, {Marino}, {Galbany}, {Bland-Hawthorn},
  {Cano-D{\'{\i}}az}, {Garc{\'{\i}}a-Benito}, {L{\'o}pez-Cob{\'a}}, {Zibetti},
  {Vilchez}, {Igl{\'e}sias-P{\'a}ramo}, {Kehrig}, {L{\'o}pez S{\'a}nchez},
  {Duarte Puertas}, \& {Ziegler}}]{2017MNRAS.469.2121S}
{S{\'a}nchez}, S.~F., {Barrera-Ballesteros}, J.~K., {S{\'a}nchez-Menguiano},
  L., {et~al.} 2017, \mnras, 469, 2121

\bibitem[{{S{\'a}nchez} {et~al.}(2012){S{\'a}nchez}, {Kennicutt}, {Gil de Paz},
  {van de Ven}, {V{\'{\i}}lchez}, {Wisotzki}, {Walcher}, {Mast}, {Aguerri},
  {Albiol-P{\'e}rez}, {Alonso-Herrero}, {Alves}, {Bakos}, {Bart{\'a}kov{\'a}},
  {Bland-Hawthorn}, {Boselli}, {Bomans}, {Castillo-Morales}, {Cortijo-Ferrero},
  {de Lorenzo-C{\'a}ceres}, {Del Olmo}, {Dettmar}, {D{\'{\i}}az}, {Ellis},
  {Falc{\'o}n-Barroso}, {Flores}, {Gallazzi}, {Garc{\'{\i}}a-Lorenzo},
  {Gonz{\'a}lez Delgado}, {Gruel}, {Haines}, {Hao}, {Husemann},
  {Igl{\'e}sias-P{\'a}ramo}, {Jahnke}, {Johnson}, {Jungwiert}, {Kalinova},
  {Kehrig}, {Kupko}, {L{\'o}pez-S{\'a}nchez}, {Lyubenova}, {Marino},
  {M{\'a}rmol-Queralt{\'o}}, {M{\'a}rquez}, {Masegosa}, {Meidt},
  {Mendez-Abreu}, {Monreal-Ibero}, {Montijo}, {Mour{\~a}o}, {Palacios-Navarro},
  {Papaderos}, {Pasquali}, {Peletier}, {P{\'e}rez}, {P{\'e}rez}, {Quirrenbach},
  {Rela{\~n}o}, {Rosales-Ortega}, {Roth}, {Ruiz-Lara},
  {S{\'a}nchez-Bl{\'a}zquez}, {Sengupta}, {Singh}, {Stanishev}, {Trager},
  {Vazdekis}, {Viironen}, {Wild}, {Zibetti}, \&
  {Ziegler}}]{2012A&A...538A...8S}
{S{\'a}nchez}, S.~F., {Kennicutt}, R.~C., {Gil de Paz}, A., {et~al.} 2012,
  \aap, 538, A8

\bibitem[{{S{\'a}nchez} {et~al.}(2013){S{\'a}nchez}, {Rosales-Ortega},
  {Jungwiert}, {Iglesias-P{\'a}ramo}, {V{\'{\i}}lchez}, {Marino}, {Walcher},
  {Husemann}, {Mast}, {Monreal-Ibero}, {Cid Fernandes}, {P{\'e}rez},
  {Gonz{\'a}lez Delgado}, {Garc{\'{\i}}a-Benito}, {Galbany}, {van de Ven},
  {Jahnke}, {Flores}, {Bland-Hawthorn}, {L{\'o}pez-S{\'a}nchez}, {Stanishev},
  {Miralles-Caballero}, {D{\'{\i}}az}, {S{\'a}nchez-Blazquez}, {Moll{\'a}},
  {Gallazzi}, {Papaderos}, {Gomes}, {Gruel}, {P{\'e}rez}, {Ruiz-Lara},
  {Florido}, {de Lorenzo-C{\'a}ceres}, {Mendez-Abreu}, {Kehrig}, {Roth},
  {Ziegler}, {Alves}, {Wisotzki}, {Kupko}, {Quirrenbach}, {Bomans}, \& {Califa
  Collaboration}}]{2013A&A...554A..58S}
{S{\'a}nchez}, S.~F., {Rosales-Ortega}, F.~F., {Jungwiert}, B., {et~al.} 2013,
  \aap, 554, A58

\bibitem[{{S{\'a}nchez-Menguiano} {et~al.}(2020){S{\'a}nchez-Menguiano},
  {S{\'a}nchez Almeida}, {Mu{\~n}oz-Tu{\~n}{\'o}n}, \&
  {S{\'a}nchez}}]{2020arXiv200914211S}
{S{\'a}nchez-Menguiano}, L., {S{\'a}nchez Almeida}, J.,
  {Mu{\~n}oz-Tu{\~n}{\'o}n}, C., \& {S{\'a}nchez}, S.~F. 2020, arXiv e-prints,
  arXiv:2009.14211

\bibitem[{{Skibba} {et~al.}(2011){Skibba}, {Engelbracht}, {Dale}, {Hinz},
  {Zibetti}, {Crocker}, {Groves}, {Hunt}, {Johnson}, {Meidt}, {Murphy},
  {Appleton}, {Armus}, {Bolatto}, {Brandl}, {Calzetti}, {Croxall}, {Galametz},
  {Gordon}, {Kennicutt}, {Koda}, {Krause}, {Montiel}, {Rix}, {Roussel},
  {Sandstrom}, {Sauvage}, {Schinnerer}, {Smith}, {Walter}, {Wilson}, \&
  {Wolfire}}]{2011ApJ...738...89S}
{Skibba}, R.~A., {Engelbracht}, C.~W., {Dale}, D., {et~al.} 2011, \apj, 738, 89

\bibitem[{{Telford} {et~al.}(2016){Telford}, {Dalcanton}, {Skillman}, \&
  {Conroy}}]{2016ApJ...827...35T}
{Telford}, O.~G., {Dalcanton}, J.~J., {Skillman}, E.~D., \& {Conroy}, C. 2016,
  \apj, 827, 35

\bibitem[{{Torrey} {et~al.}(2018){Torrey}, {Vogelsberger}, {Hernquist},
  {McKinnon}, {Marinacci}, {Simcoe}, {Springel}, {Pillepich}, {Naiman},
  {Pakmor}, {Weinberger}, {Nelson}, \& {Genel}}]{2018MNRAS.477L..16T}
{Torrey}, P., {Vogelsberger}, M., {Hernquist}, L., {et~al.} 2018, \mnras, 477,
  L16

\bibitem[{{Tremonti} {et~al.}(2004){Tremonti}, {Heckman}, {Kauffmann},
  {Brinchmann}, {Charlot}, {White}, {Seibert}, {Peng}, {Schlegel}, {Uomoto},
  {Fukugita}, \& {Brinkmann}}]{2004ApJ...613..898T}
{Tremonti}, C.~A., {Heckman}, T.~M., {Kauffmann}, G., {et~al.} 2004, \apj, 613,
  898

\bibitem[{{Tsamis} {et~al.}(2011){Tsamis}, {Walsh}, {V{\'\i}lchez}, \&
  {P{\'e}quignot}}]{2011MNRAS.412.1367T}
{Tsamis}, Y.~G., {Walsh}, J.~R., {V{\'\i}lchez}, J.~M., \& {P{\'e}quignot}, D.
  2011, \mnras, 412, 1367

\bibitem[{{Van Der Walt} {et~al.}(2011){Van Der Walt}, {Colbert}, \&
  {Varoquaux}}]{2011arXiv1102.1523V}
{Van Der Walt}, S., {Colbert}, S.~C., \& {Varoquaux}, G. 2011, ArXiv e-prints

\bibitem[{{Vila-Costas} \& {Edmunds}(1992)}]{1992MNRAS.259..121V}
{Vila-Costas}, M.~B. \& {Edmunds}, M.~G. 1992, \mnras, 259, 121

\bibitem[{{V{\'\i}lchez} {et~al.}(2019){V{\'\i}lchez}, {Rela{\~n}o},
  {Kennicutt}, {De Looze}, {Moll{\'a}}, \& {Galametz}}]{2019MNRAS.483.4968V}
{V{\'\i}lchez}, J.~M., {Rela{\~n}o}, M., {Kennicutt}, R., {et~al.} 2019,
  \mnras, 483, 4968

\bibitem[{{Yates} {et~al.}(2012){Yates}, {Kauffmann}, \&
  {Guo}}]{2012MNRAS.422..215Y}
{Yates}, R.~M., {Kauffmann}, G., \& {Guo}, Q. 2012, \mnras, 422, 215

\bibitem[{{Yin} {et~al.}(2009){Yin}, {Hou}, {Prantzos}, {Boissier}, {Chang},
  {Shen}, \& {Zhang}}]{2009A&A...505..497Y}
{Yin}, J., {Hou}, J.~L., {Prantzos}, N., {et~al.} 2009, \aap, 505, 497

\bibitem[{{Zahid} {et~al.}(2014){Zahid}, {Dima}, {Kudritzki}, {Kewley},
  {Geller}, {Hwang}, {Silverman}, \& {Kashino}}]{2014ApJ...791..130Z}
{Zahid}, H.~J., {Dima}, G.~I., {Kudritzki}, R.-P., {et~al.} 2014, \apj, 791,
  130

\bibitem[{{Zurita} \& {Bresolin}(2012)}]{2012MNRAS.427.1463Z}
{Zurita}, A. \& {Bresolin}, F. 2012, \mnras, 427, 1463

\end{thebibliography}

%__________________________________________________________________
%% Appendix Supplementary material %%%%%%%%%%%%%%%%%%%%%%
\begin{appendix}
\section{Supplementary material}
\label{sec:sm}

In this appendix the O/H -- SFR relation as a function of the M$_\star$ and the parameter D$\rm_n$ (4000) for the galaxies in Fig.~\ref{fig:3_OHNOSFR_d4000} is shown in detail. 
From the O/H -- SFR relation, a subsample of 21 galaxies is randomly defined in selected zones below the {\it isochrone} D$\rm_n$(4000) $\leq$ 1.2. Its location on the SFR -- M$_\star$ diagram and its spectra and SDSS three-colour images are shown. An upcoming more detailed study on the characterisation of this subsample of galaxies will be presented in a forthcoming work (Duarte Puertas et al. in prep.).

\begin{figure}
    \centering
	\includegraphics[width=\columnwidth]{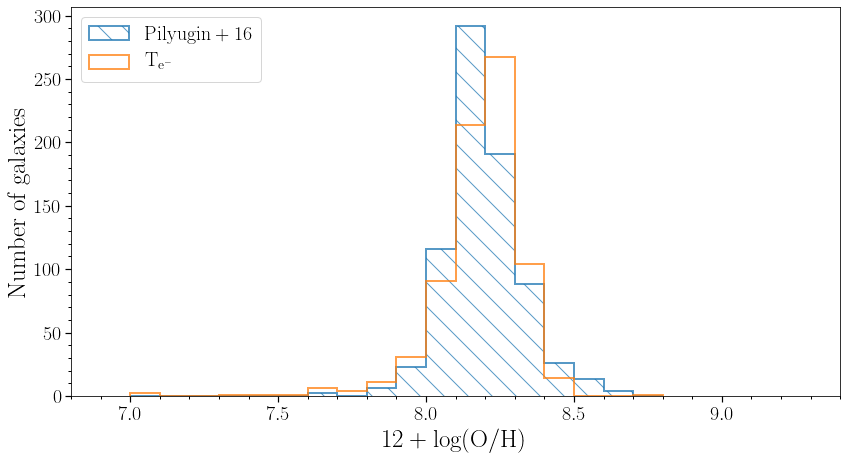}	
	\caption{Distribution of oxygen abundance for SDSS star-forming galaxies with S/N$\rm ([OIII]\lambda4363) > 5$ using Pi16 methodology (blue dashed histogram) and using direct (electron temperature based) methodology (orange open histogram).}
\label{fig:A5_OH_te}
\end{figure}

We have checked the SFR derived for these galaxies, confirming that these values were not produced by aperture corrections since the correction applied to these outliers was in fact smaller than average. We have searched and extracted the fluxes of the measured SDSS spectra and have verified for all these galaxies with electron temperature (i.e. with $\rm S/N([OIII]\lambda4363) > 5$) that their direct (electron temperature based) abundances agree with our oxygen abundance values to within $\sim$-0.025 dex on average (see Fig.~\ref{fig:A5_OH_te}), as expected.

% Figure Ap 1, 9__________________________________________________________________
%__________________________________________________________________

\begin{figure*}
    \centering
    \includegraphics[width=.9\textwidth]{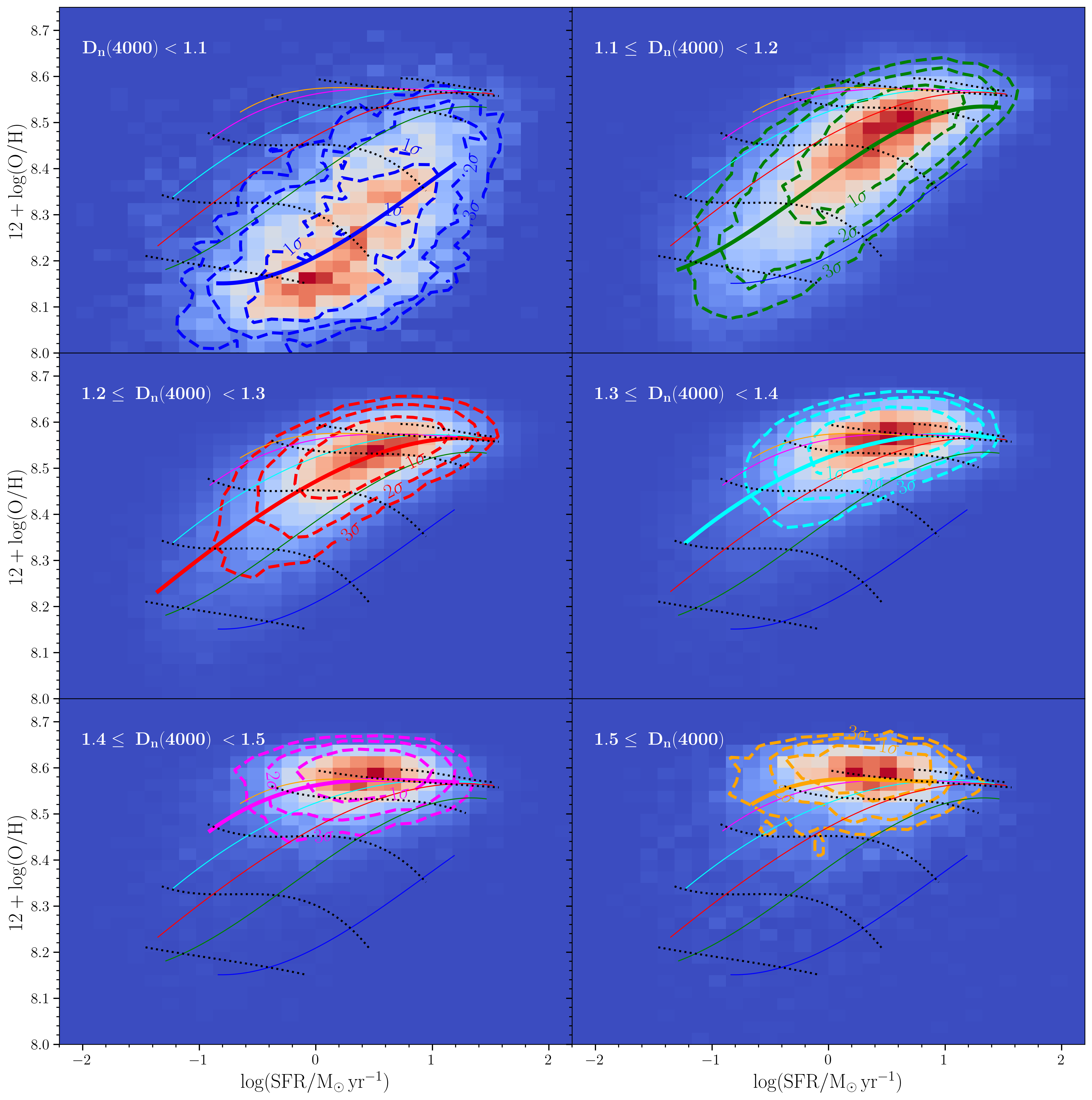}
    \caption{Detailed analysis of the 12+log(O/H) -- SFR density plot presenting confidence limits for each D$\rm _n$(4000) bin. Density plots for the relation between the 12+log(O/H) and SFR for star-forming galaxies for six D$\rm _n$(4000) ranges. All the lines have the same colours as in Fig.~\ref{fig:3_OHNOSFR_d4000}. The dashed lines represent the 1$\sigma$, 2$\sigma$, and 3$\sigma$ contours in each D$\rm _n$(4000) bin.}
    \label{fig:A1_MZRSFR_binn_fit}
\end{figure*}

Fig.~\ref{fig:A1_MZRSFR_binn_fit} shows a density plot of our aperture corrected O/H versus SFR in six ranges of D$\rm_n$(4000). In each range of D$\rm_n$(4000), the density plot of the O/H -- SFR relation of the galaxies in that interval is shown and the contours 1$\sigma$, 2$\sigma$ and 3$\sigma$ in each bin are represented, highlighting the fit to the running median.

% Figure Ap 2, 10__________________________________________________________________
%__________________________________________________________________
%\clearpage

\begin{figure*}
\centering
%\begin{minipage}{.48\linewidth}
    \includegraphics[width=.48\textwidth]{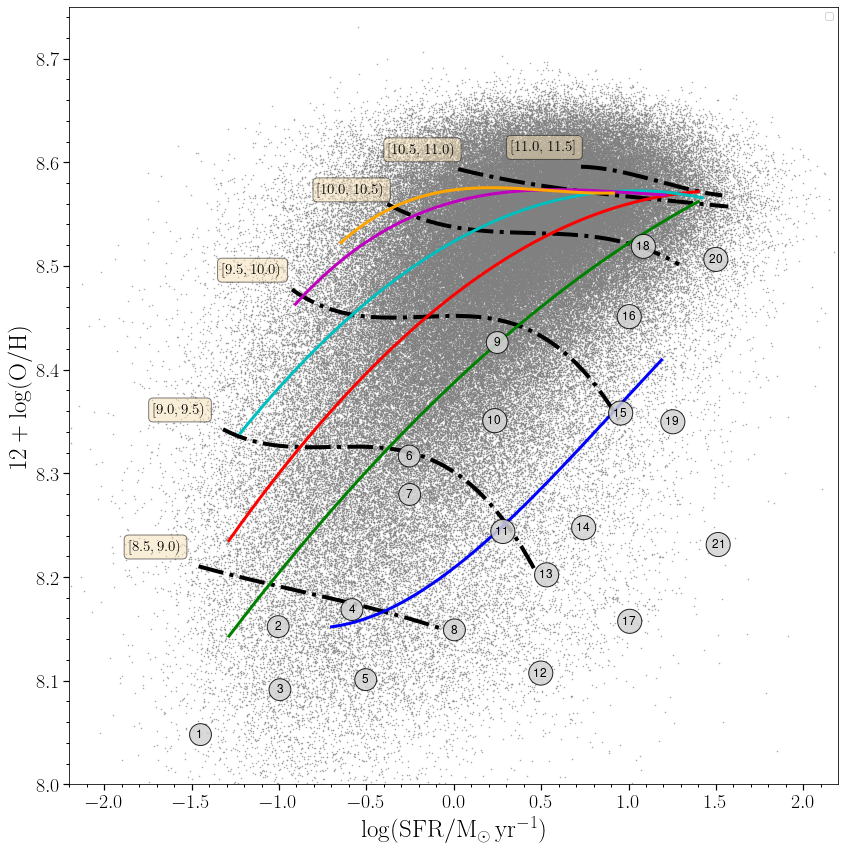}
%\end{minipage}
\hfill
%\begin{minipage}{.48\linewidth}
    \includegraphics[width=.48\textwidth]{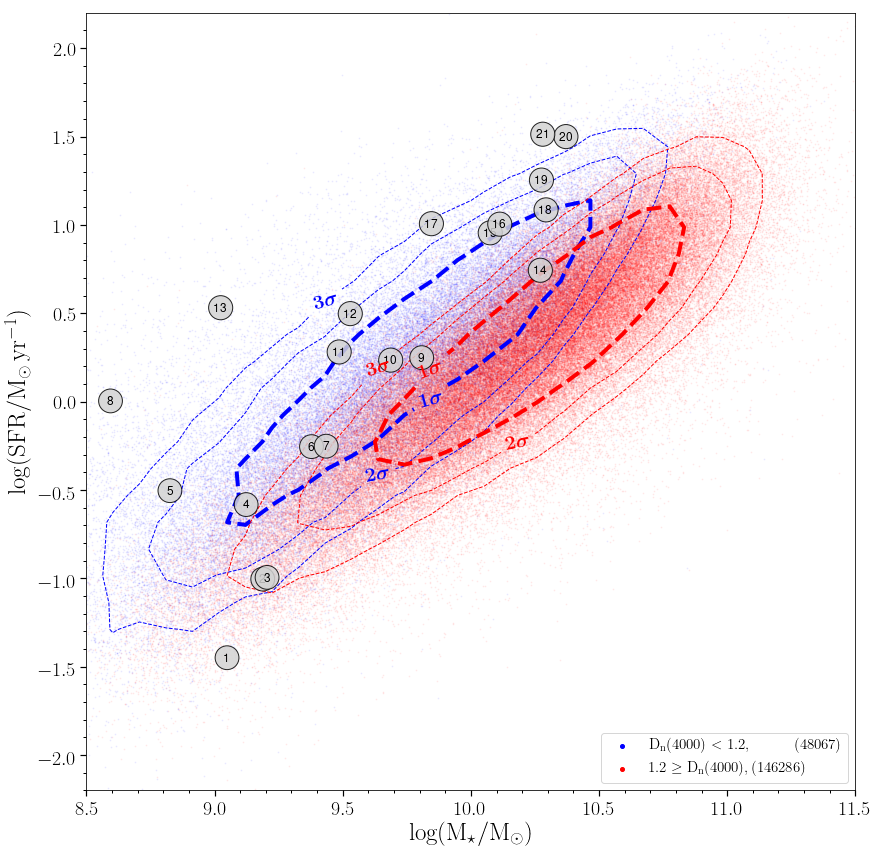}
%\end{minipage}
\caption{Selection of 21 galaxies from the 12+log(O/H) vs. SFR diagram. Left panel) Relation between 12+log(O/H) and SFR for star-forming galaxies. All the lines have the same colours as Fig.~\ref{fig:3_OHNOSFR_d4000}. A sample of 21 galaxies has been selected (as an example) in this diagram, the positions are represented with grey circles and have been labelled from 1 to 21 according to their SFR. Right panel) Relation between SFR and M$_\star$ for star-forming galaxies. The points have been colour coded according to their D$\rm _n$(4000), blue if D$\rm _n$(4000) $<$ 1.2 and red if D$\rm _n$(4000) $\geq$ 1.2. The dashed lines represent the 1$\sigma$, 2$\sigma$, and 3$\sigma$ contours in each D$\rm _n$(4000) sub-sample. Grey circles shows the position in the SFR-M$_\star$ diagram of the 21 galaxies.}
\label{fig:A2_sSFRSFRM_d4000}
\end{figure*}

Fig.~\ref{fig:A2_sSFRSFRM_d4000} shows again the aperture corrected O/H -- SFR relation (left panel) and the aperture corrected SFR -- M$_\star$ relation (right panel). Since a population of galaxies has been found in the zone where O/H dilution occurs, we have defined there a subsample of 21 galaxies randomly selected, distributed below the {\it isochrone} 1.1 $<$ D$\rm_n$(4000) $\leq$ 1.2 in the O/H -- SFR diagram. We have done the selection in that way that some of them must be found in the {\it isochrones} of D$\rm_n$(4000) < 1.1 and others in 1.1 $<$ D$\rm_n$(4000) $\leq $ 1.2, as well as in the iso-mass [9, 9.5]. The positions of the 21 galaxies, as well as the position of the subsample of galaxies that have values of D$\rm_n$(4000) $ < $ 1.2 (blue color) and that of D$\rm_n$(4000) $\geq$ 1.2 (red color), are also shown in the SFR -- M$_\star$ relation. Galaxies with values of D$\rm_n$(4000) $ < $ 1.2 (found in the lower right of the O/H -- SFR diagram) are located in the upper left of the SFR -- M$_\star$ diagram. As expected from Fig.~\ref{fig:3_OHNOSFR_d4000}, galaxies with younger stellar populations have higher SFR values than older ones at a given stellar mass.

% Figure Ap 3, 11__________________________________________________________________
%__________________________________________________________________

As said in the main text, in order to explore the nature of this subsample of 8598 galaxies, we have performed a sanity check to confirm their metallicity and SFR, such as we show in Figures~\ref{fig:A2_sSFRSFRM_d4000}, \ref{fig:A3_spectra}, and \ref{fig:A4_image}, to gain more insight into the nature of this sub-sample of galaxies.
In Figs.~\ref{fig:A3_spectra} and \ref{fig:A4_image} the spectra and SDSS three-colour images of 21 selected galaxies are shown. By focusing on the spectra of the galaxies found in the iso-mass [9, 9.5] (galaxies 6, 11, and 13) it can be observed how excitation decreases as the age of the galaxy's stellar populations, and 12+log(O/H) increases. In the {\it isochrone} D$\rm_n$(4000) < 1.1 the spectra of these three galaxies (galaxies 4, 11, and 15) look similar, while their morphologies are different.

% Figure Ap 4, 12__________________________________________________________________
\begin{figure*}
    \centering
    \includegraphics[width=\textwidth]{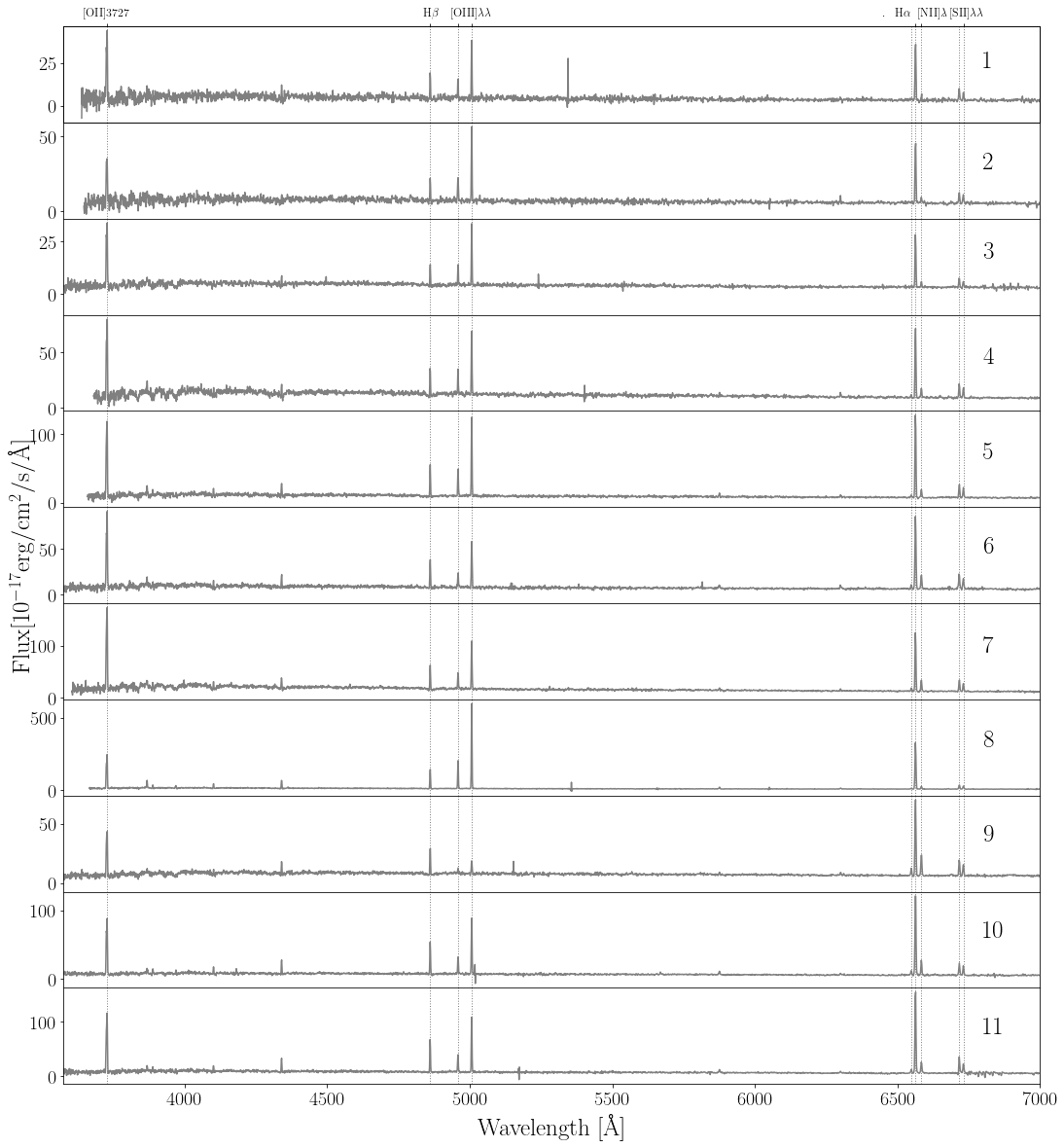}
    \caption{Spectra of the 21 galaxies defined in Fig.~\ref{fig:A2_sSFRSFRM_d4000}. Grey dotted lines show the location of the emission lines studied in this paper. The galaxy label is indicated in each panel (right part).}
    \label{fig:A3_spectra}
\end{figure*}
%__________________________________________________________________
\addtocounter{figure}{-1}
%__________________________________________________________________
\begin{figure*}
    \centering
    \includegraphics[width=\textwidth]{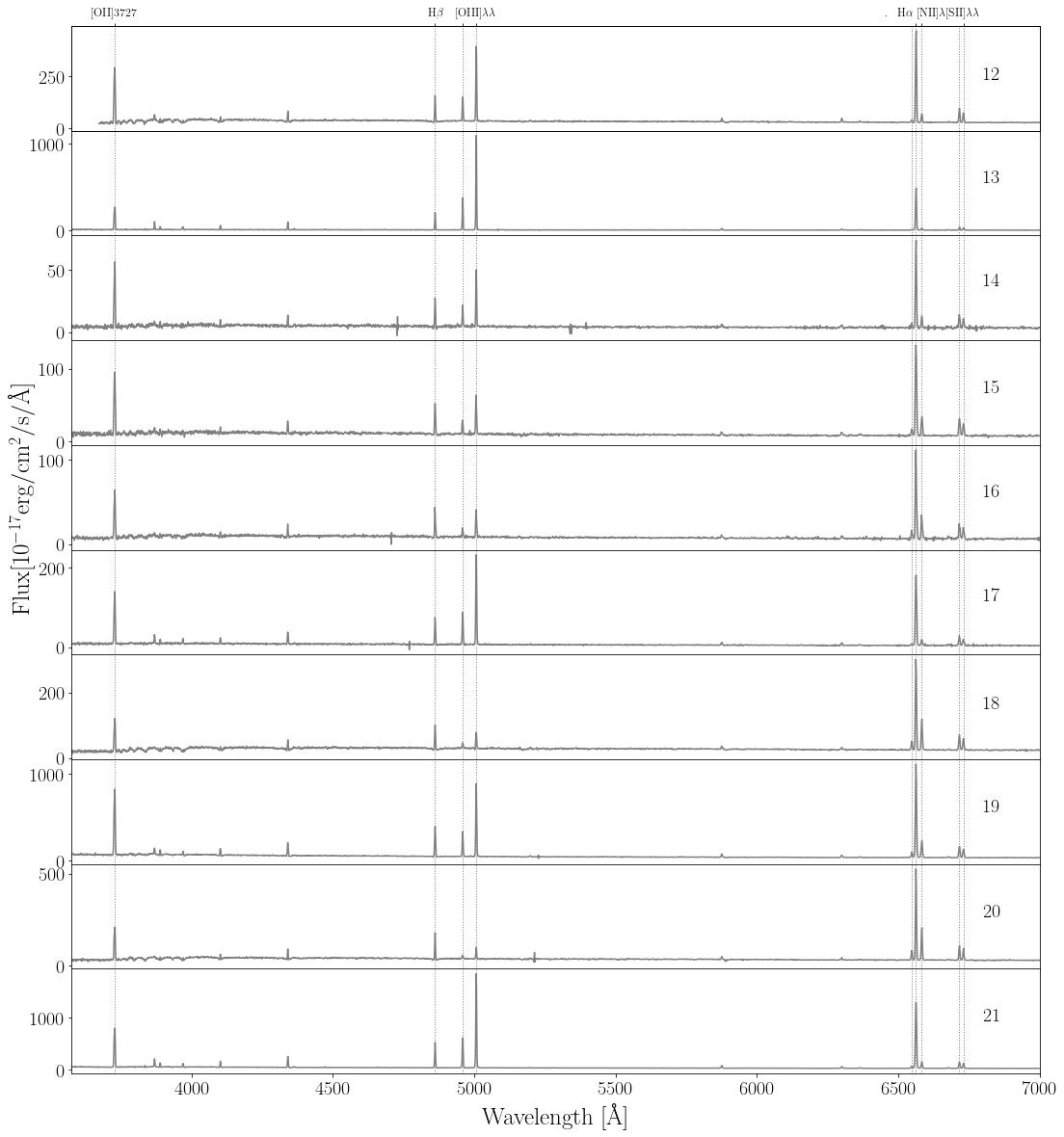}
    \caption{continued.}
\end{figure*}
%__________________________________________________________________

% Figure Ap 5, 13__________________________________________________________________
\begin{figure*}
    \centering
    \includegraphics[width=\textwidth]{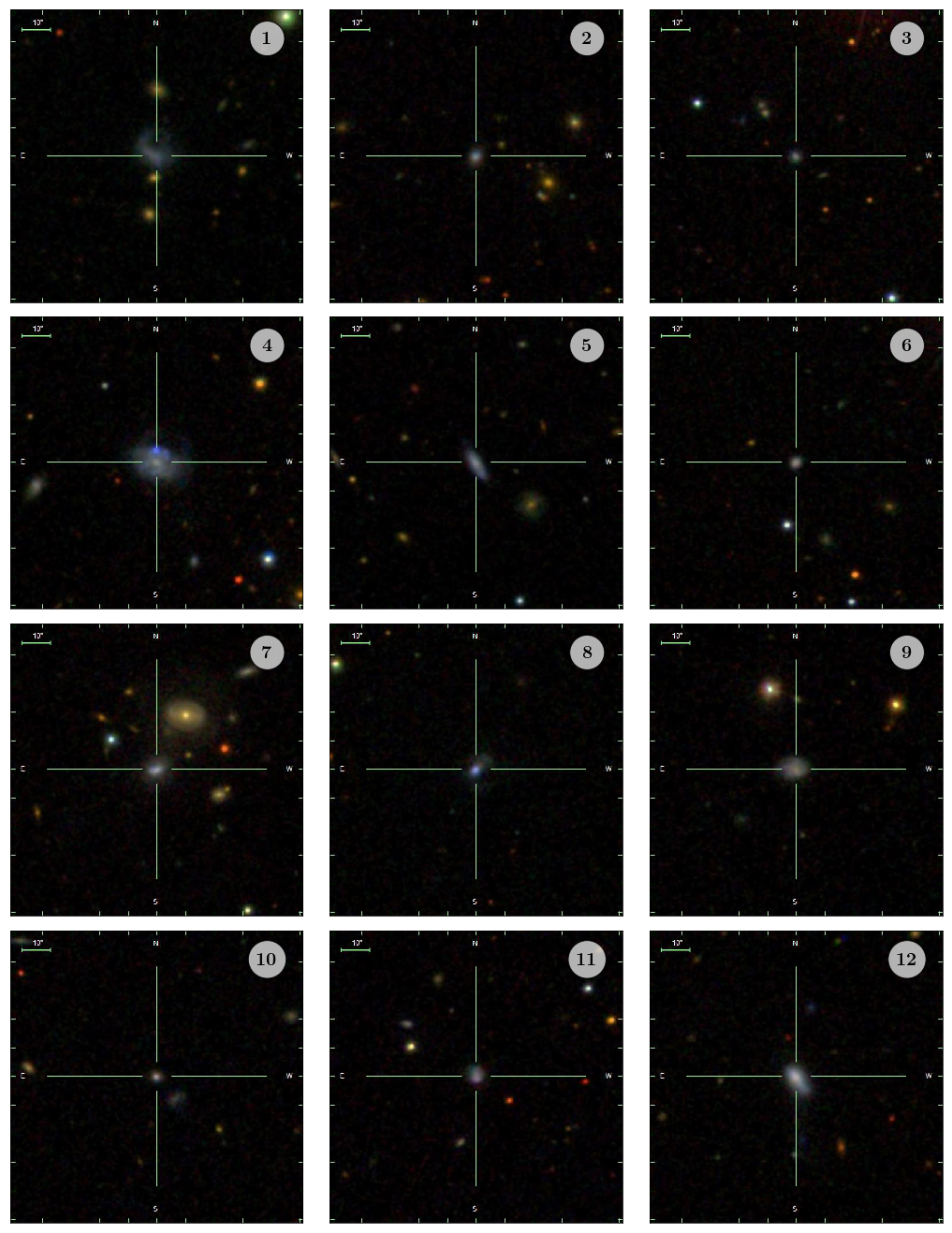}
    \caption{SDSS three-colour images of 21 galaxies defined in Fig.~\ref{fig:A2_sSFRSFRM_d4000} in the SDSS-DR12 footprint. North is up and East is left. The galaxy label is indicated in each panel (upper right).}
    \label{fig:A4_image}
\end{figure*}
%__________________________________________________________________
\addtocounter{figure}{-1}
%__________________________________________________________________
\begin{figure*}
    \centering
    \includegraphics[width=\textwidth]{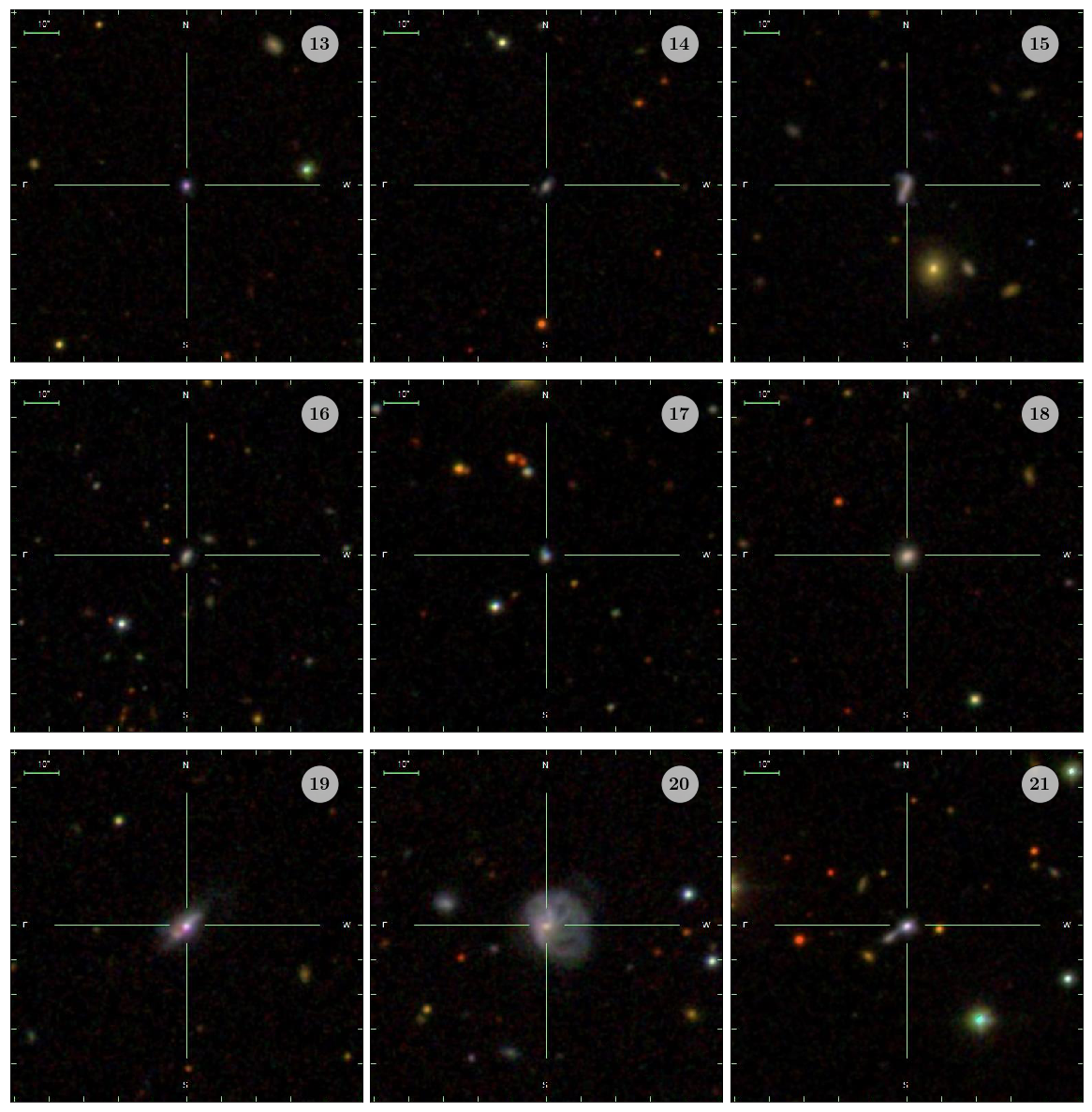}
    \caption{continued.}
\end{figure*}
%__________________________________________________________________

\end{appendix}
\end{document}